\pdfoutput=1
\documentclass[12pt,a4paper]{article}

\usepackage{lineno} 
\usepackage{rotating}
\usepackage{graphicx}  
\usepackage{xspace} 
                    
\usepackage{color}
\usepackage{colortbl}
\usepackage{amsmath} 
\usepackage{ifthen} 

\usepackage{multirow}
\usepackage{verbatim}

\newboolean{pdflatex}
\setboolean{pdflatex}{true} 
\newboolean{articletitles}
\setboolean{articletitles}{true} 

\newboolean{uprightparticles}
\setboolean{uprightparticles}{false} 

\usepackage{amssymb}
\usepackage{amsfonts}
\usepackage{upgreek} 
\usepackage{subfig}

\usepackage{hyperref}  
\usepackage[all]{hypcap}

\def\lhcb {LHCb\xspace}
\def\ux85 {UX85\xspace}

\ifthenelse{\boolean{uprightparticles}}%
{

 \def\PDelta      {\ensuremath{\Delta}\xspace}
 \def\PXi      {\ensuremath{\Xi}\xspace}
 \def\PLambda      {\ensuremath{\Lambda}\xspace}
 \def\PSigma      {\ensuremath{\Sigma}\xspace}
 \def\POmega      {\ensuremath{\Omega}\xspace}
 \def\PUpsilon      {\ensuremath{\Upsilon}\xspace}

 \def\PB      {\ensuremath{\mathrm{B}}\xspace}
 
 \def\PD      {\ensuremath{\mathrm{D}}\xspace}

 \def\PK      {\ensuremath{\mathrm{K}}\xspace}

 \def\Pb      {\ensuremath{\mathrm{b}}\xspace}
 \def\Pc      {\ensuremath{\mathrm{c}}\xspace}

 \def\Pi      {\ensuremath{\mathrm{i}}\xspace}

}
{

 \mathchardef\PDelta="7101
 \mathchardef\PXi="7104
 \mathchardef\PLambda="7103
 \mathchardef\PSigma="7106
 \mathchardef\POmega="710A
 \mathchardef\PUpsilon="7107
 
 \def\PB      {\ensuremath{B}\xspace}
 
 \def\PD      {\ensuremath{D}\xspace}

 \def\PK      {\ensuremath{K}\xspace}

 \def\Pb      {\ensuremath{b}\xspace}
 \def\Pc      {\ensuremath{c}\xspace}

 \def\Pi      {\ensuremath{i}\xspace}

}

\def\cquark    {\ensuremath{\Pc}\xspace}

\def\bquark    {\ensuremath{\Pb}\xspace}

\def\kaon  {\ensuremath{\PK}\xspace}
  \def\Kbar  {\kern 0.2em\overline{\kern -0.2em \PK}{}\xspace}

\def\Kz    {\ensuremath{\kaon^0}\xspace}
\def\Kzb   {\ensuremath{\Kbar^0}\xspace}
\def\KzKzb {\ensuremath{\Kz \kern -0.16em \Kzb}\xspace}
\def\Kp    {\ensuremath{\kaon^+}\xspace}
\def\Km    {\ensuremath{\kaon^-}\xspace}

\def\KpKm  {\ensuremath{\Kp \kern -0.16em \Km}\xspace}

  \def\Dbar    {\kern 0.2em\overline{\kern -0.2em \PD}{}\xspace}
\def\D       {\ensuremath{\PD}\xspace}

\def\Dz      {\ensuremath{\D^0}\xspace}
\def\Dzb     {\ensuremath{\Dbar^0}\xspace}
\def\DzDzb   {\ensuremath{\Dz {\kern -0.16em \Dzb}}\xspace}
\def\Dp      {\ensuremath{\D^+}\xspace}
\def\Dm      {\ensuremath{\D^-}\xspace}

\def\DpDm    {\ensuremath{\Dp {\kern -0.16em \Dm}}\xspace}

\def\Ds      {\ensuremath{\D^+_\squark}\xspace}

  \def\Bbar    {\kern 0.18em\overline{\kern -0.18em \PB}{}\xspace}

  \def\Y#1S{\ensuremath{\PUpsilon{(#1S)}}\xspace}

\def\to                 {\ensuremath{\rightarrow}\xspace}

\def\CP                {\ensuremath{C\!P}\xspace}

\def\AT#1     {\ensuremath{A_{\mathrm{T}}^{#1}}\xspace}           

\def\C#1      {\ensuremath{\mathcal{C}_{#1}}\xspace}                       
\def\Cp#1     {\ensuremath{\mathcal{C}_{#1}^{'}}\xspace}                    
\def\Ceff#1   {\ensuremath{\mathcal{C}_{#1}^{\mathrm{(eff)}}}\xspace}        
\def\Cpeff#1  {\ensuremath{\mathcal{C}_{#1}^{'\mathrm{(eff)}}}\xspace}       
\def\Ope#1    {\ensuremath{\mathcal{O}_{#1}}\xspace}                       
\def\Opep#1   {\ensuremath{\mathcal{O}_{#1}^{'}}\xspace}                    

\newcommand{\tev}{\ensuremath{\mathrm{\,Te\kern -0.1em V}}\xspace}
\newcommand{\gev}{\ensuremath{\mathrm{\,Ge\kern -0.1em V}}\xspace}
\newcommand{\mev}{\ensuremath{\mathrm{\,Me\kern -0.1em V}}\xspace}
\newcommand{\kev}{\ensuremath{\mathrm{\,ke\kern -0.1em V}}\xspace}
\newcommand{\ev}{\ensuremath{\mathrm{\,e\kern -0.1em V}}\xspace}
\newcommand{\gevc}{\ensuremath{{\mathrm{\,Ge\kern -0.1em V\!/}c}}\xspace}
\newcommand{\mevc}{\ensuremath{{\mathrm{\,Me\kern -0.1em V\!/}c}}\xspace}
\newcommand{\gevcc}{\ensuremath{{\mathrm{\,Ge\kern -0.1em V\!/}c^2}}\xspace}
\newcommand{\gevgevcccc}{\ensuremath{{\mathrm{\,Ge\kern -0.1em V^2\!/}c^4}}\xspace}
\newcommand{\mevcc}{\ensuremath{{\mathrm{\,Me\kern -0.1em V\!/}c^2}}\xspace}

\def\mum  {\ensuremath{\,\upmu\rm m}\xspace}

\def\fb   {\ensuremath{\mbox{\,fb}}\xspace}
\def\invfb   {\ensuremath{\mbox{\,fb}^{-1}}\xspace}

\def\fs   {\ensuremath{\rm \,fs}\xspace}

\def\gsim{{~\raise.15em\hbox{$>$}\kern-.85em
          \lower.35em\hbox{$\sim$}~}\xspace}
\def\lsim{{~\raise.15em\hbox{$<$}\kern-.85em
          \lower.35em\hbox{$\sim$}~}\xspace}

\def\rad{\ensuremath{\rm \,rad}\xspace}

\def\evtgen     {\mbox{\textsc{EvtGen}}\xspace}
\def\photos     {\mbox{\textsc{Photos}}\xspace}
\def\pythia     {\mbox{\textsc{Pythia}}\xspace}

\def\geant      {\mbox{\textsc{Geant4}}\xspace}

\def\tell1  {TELL1\xspace}
\def\ukl1   {UKL1\xspace}

\usepackage{mciteplus}

\begin{document}

\newcommand{\BdDecay}{\texorpdfstring{$B^0 \to J/\psi K^{*}(892)^0$ }{Bd}}
\newcommand{\BsDecay}{\texorpdfstring{$B_s^0 \to J/\psi \phi$ }{Bs}}
\newcommand{\Apara}{\texorpdfstring{$A_{\parallel}$ }{Apara}}
\newcommand{\Aperp}{\texorpdfstring{$A_{\perp}$ }{Aperp}}
\newcommand{\Azero}{\texorpdfstring{$A_{0}$ }{Azero}}
\newcommand{\As}{\texorpdfstring{$A_{\rm S}$ }{As}}

\def\Aparasq{|A_{\parallel}|^2}
\def\Aperpsq{|A_{\perp}|^2 }
\def\Azerosq{|A_{0}|^2 }
\def\Assq{|A_{\mathrm{S}}|^2 }
\def\Fs{F_{\mathrm{S}} }

\def\Dpara{\delta_{\parallel} }
\def\Dperp{\delta_{\perp} }
\def\Dzero{\delta_{0} }
\def\Ds{\delta_{\mathrm{S}} }

\def\Aparasqbar{|\overline{A}_{\parallel}|^2}
\def\Aperpsqbar{|\overline{A}_{\perp}|^2 }
\def\Azerosqbar{|\overline{A}_{0}|^2 }
\def\Assqbar{|\overline{A}_{\mathrm{S}}|^2 }
\def\Fsbar{\overline{F}_{\mathrm{S}} }

\def\Dparabar{\overline{\delta}_{\parallel} }
\def\Dperpbar{\overline{\delta}_{\perp} }
\def\Dzerobar{\overline{\delta}_{0} }
\def\Dsbar{\overline{\delta}_{\mathrm{S}} }

\newcommand{\Jpsi}{\texorpdfstring{J /\psi }{Jpsi}}


\begin{titlepage}
\pagenumbering{roman}

\vspace*{-1.5cm}
\centerline{\large EUROPEAN ORGANIZATION FOR NUCLEAR RESEARCH (CERN)}
\vspace*{1.5cm}
\hspace*{-0.5cm}
\begin{tabular*}{\linewidth}{lc@{\extracolsep{\fill}}r}
\ifthenelse{\boolean{pdflatex}}
{\vspace*{-2.7cm}\mbox{\!\!\!\includegraphics[width=.14\textwidth]{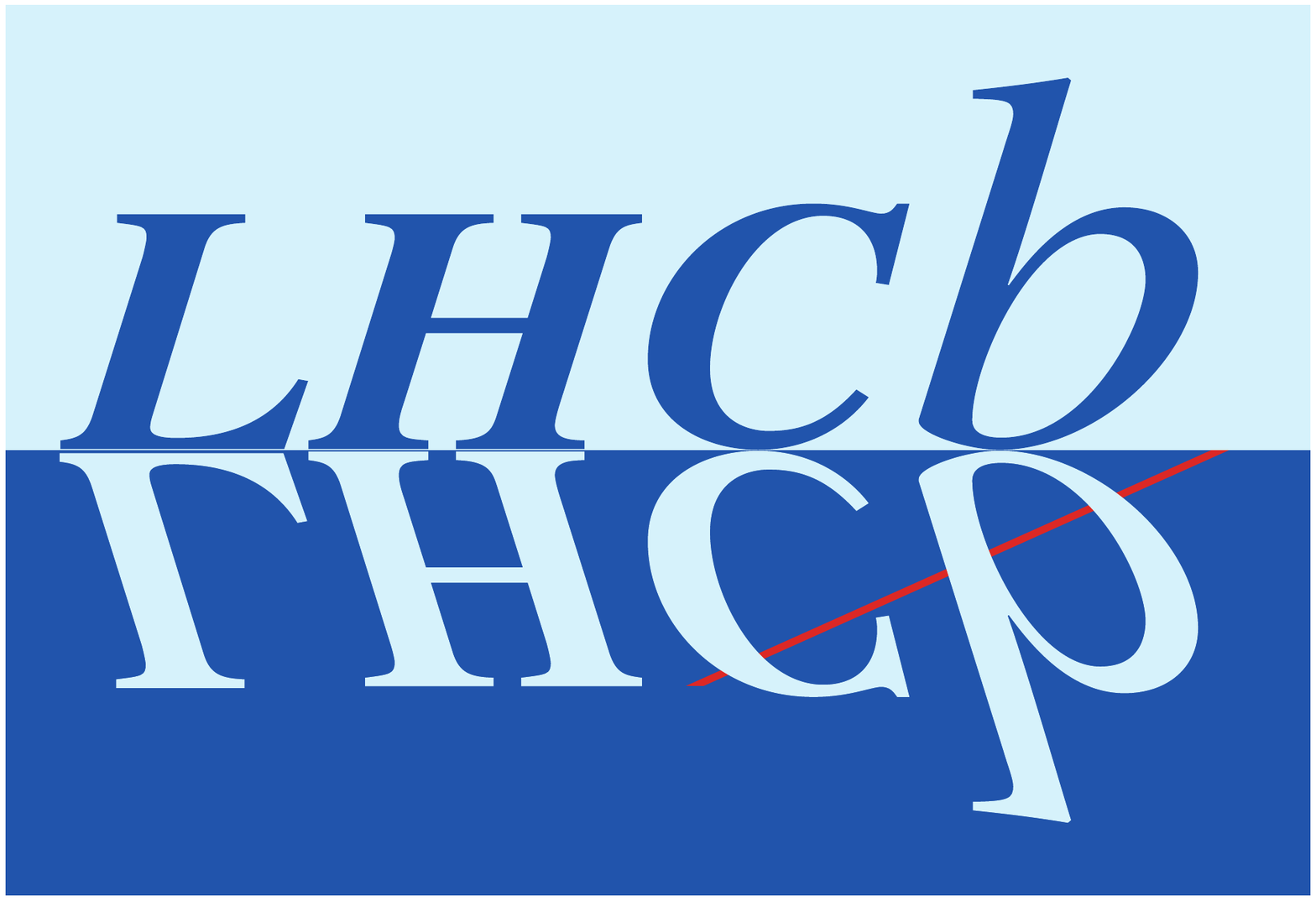}} & &}%
{\vspace*{-1.2cm}\mbox{\!\!\!\includegraphics[width=.12\textwidth]{lhcb-logo.eps}} & &}%
\\
 & & CERN-PH-EP-2013-104 \\  
 & & LHCb-PAPER-2013-023 \\  
 & & September 5, 2013 \\
 & & \\

\end{tabular*}

\vspace*{1.5cm}

{\bf\boldmath\huge
\begin{center}
Measurement of the polarisation amplitudes in $B^0 \to J/\psi K^{*}(892)^0$ decays 
\end{center}
}

\vspace*{1.5cm}

\begin{center}
The LHCb collaboration\footnote{Authors are listed on the following pages.}
\end{center}

\vspace{\fill}

\begin{abstract}
  \noindent
An  analysis of the decay $B^0 \to J/\psi K^{*}(892)^0$ is presented 
using data, corresponding to an integrated luminosity of $1.0$\invfb, collected in 
$pp$ collisions at a centre-of-mass energy of $7 \tev$ with the LHCb detector. 
The polarisation amplitudes and the corresponding phases are measured to be 
      \[
  \setlength{\arraycolsep}{1mm}
  \begin{array}{cclllllll}
  \Aparasq &\;=\; & 0.227  &\pm & 0.004  & \text{(stat.)} &\pm & 0.011 & \text{(syst.)}, \\
  \Aperpsq  &\;=\; & 0.201   &\pm & 0.004    & \text{(stat.)} &\pm & 0.008&\text{(syst.)},\\
  \Dpara\; \text{[rad]} &\;=\; & -2.94 &\pm & 0.02 &\text{(stat.)} &\pm &0.03&\text{(syst.)},\\  
  \Dperp\; \text{[rad]} &\;=\; & \phantom{-}2.94 &\pm & 0.02 &\text{(stat.)} &\pm &0.02&\text{(syst.)}.\\
  \end{array}
  \]
Comparing $B^0\to J/\psi K^{*}(892){^0}$ and $\overline{B}^0\to J/\psi \overline K^{*}(892){^0}$ decays, no evidence 
for direct \CP violation is found.
\end{abstract}

\vspace*{1.0cm}

\begin{center}
Submitted to Phys. Rev. D 
\end{center}

\vspace{\fill}

\begin{center}
\small{\copyright CERN on behalf of the LHCb collaboration, license
\href{http://creativecommons.org/licenses/by/3.0/}{CC-BY-3.0}}
\end{center}

\end{titlepage}


\newpage
\setcounter{page}{2}
\mbox{~}
\newpage

\centerline{\large\bf LHCb collaboration}
\begin{flushleft}
\small
R.~Aaij$^{40}$, 
C.~Abellan~Beteta$^{35,n}$, 
B.~Adeva$^{36}$, 
M.~Adinolfi$^{45}$, 
C.~Adrover$^{6}$, 
A.~Affolder$^{51}$, 
Z.~Ajaltouni$^{5}$, 
J.~Albrecht$^{9}$, 
F.~Alessio$^{37}$, 
M.~Alexander$^{50}$
% ...

{\footnotesize \it
% ...
$ ^{5}$Clermont Universit\'{e}, Universit\'{e} Blaise Pascal, CNRS/IN2P3, LPC, Clermont-Ferrand, France\\
$ ^{6}$CPPM, Aix-Marseille Universit\'{e}, CNRS/IN2P3, Marseille, France\\
% ...
$ ^{9}$Fakult\"{a}t Physik, Technische Universit\"{a}t Dortmund, Dortmund, Germany\\
% ...
$ ^{35}$Universitat de Barcelona, Barcelona, Spain\\
$ ^{36}$Universidad de Santiago de Compostela, Santiago de Compostela, Spain\\
$ ^{37}$European Organization for Nuclear Research (CERN), Geneva, Switzerland\\
% ...
$ ^{40}$Nikhef National Institute for Subatomic Physics, Amsterdam, The Netherlands\\
% ...
$ ^{45}$H.H. Wills Physics Laboratory, University of Bristol, Bristol, United Kingdom\\
% ...
$ ^{50}$School of Physics and Astronomy, University of Glasgow, Glasgow, United Kingdom\\
$ ^{51}$Oliver Lodge Laboratory, University of Liverpool, Liverpool, United Kingdom\\
% ...
$ ^{n}$LIFAELS, La Salle, Universitat Ramon Llull, Barcelona, Spain\\
% ...
}
\end{flushleft}

\cleardoublepage




\pagestyle{plain} %
\setcounter{page}{1}
\pagenumbering{arabic}


\clearpage
\newpage
\section{Introduction}
\label{sec:Introduction}
The measurement of the polarisation content of the
decay \mbox{$B^0\to J/\psi(\mu^+\mu^-) K^{*}{^0}(K^+\pi^-)$} and its
charge-conjugate \mbox{$\overline{B}^0\to J/\psi(\mu^+\mu^-) \overline K^{*}{^0}(K^-\pi^+)$}
is presented in this paper, where the notation $K^{*}{^0}$ is used to refer
to the $K^{*}(892){^0}$ meson. Recent measurements have been performed by
BaBar (2007,~\cite{Collaboration:2007kx}), Belle (2005,~\cite{PhysRevLett.95.091601}) and CDF (2005,~\cite{cdf}).
A detailed comparison can be found in Sec.~\ref{sec:results}.
The decay can be decomposed in terms of three transversity states,
corresponding to the relative orientation of the linear polarisation vectors of
the two vector mesons.  The amplitudes are referred to as P-wave amplitudes since the
$K\pi$ system is in a P-wave state, and are denoted by $A_0$
(longitudinal),  $A_\parallel$ (transverse-parallel) and $A_\perp$
(transverse-perpendicular),
where the relative orientations are shown in parentheses. 
An additional S-wave amplitude corresponding to a non-resonant  $K\pi$ system is denoted by $A_{\rm S}$.
The strong phases of the four amplitudes are $\Dzero$, $\Dpara$, $\Dperp$ and $\Ds$, respectively, and by convention 
$\Dzero$ is set to zero.
The parity of the final states is even for $A_0$ and $A_\parallel$, and odd for $A_\perp$ and $A_{\rm S}$.

\begin{figure}[!h]
\begin{center}
\subfloat[Tree]{
\includegraphics[width=7cm]{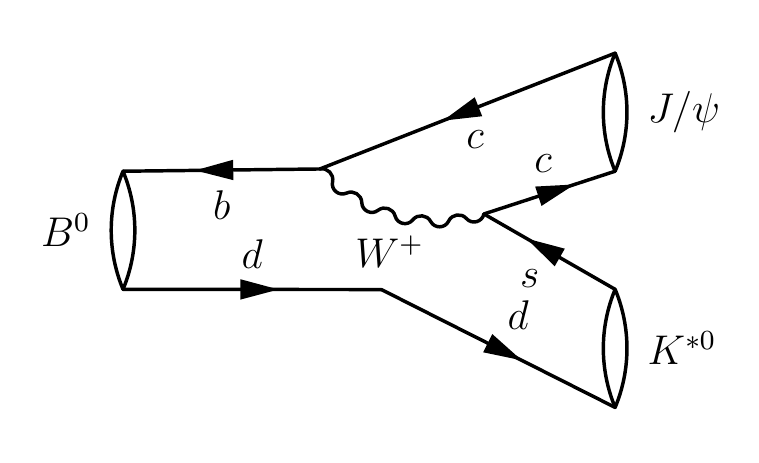}
\label{fig:tree}}
\subfloat[Penguin]{
\includegraphics[width=7cm]{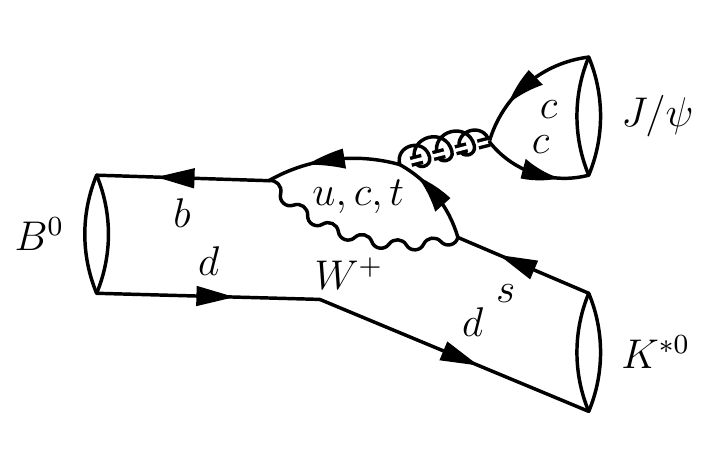}
\label{fig:gluPen}}
\caption{\small Feynman diagrams contributing to $B^0 \to J/\psi K^{*0}$ decays.}
\label{fig:treePen}
\end{center}
\end{figure}

The Standard Model (SM) predicts that the \mbox{$B^0\to J/\psi(\mu^+\mu^-) K^{*}{^0}(K^+\pi^-)$} decay is dominated by a
colour-suppressed tree diagram (Fig.~\ref{fig:tree}), with highly-suppressed
contributions from gluonic and electroweak loop (penguin) diagrams
(Fig.~\ref{fig:gluPen}). Neglecting the penguin contributions and using na\"ive
factorisation for the tree diagram leads to predictions for the P-wave
amplitudes \mbox{$|A_0|^2\approx 0.5$},  and $A_\parallel\approx
A_\perp$~\cite{Aleksan:1994bh}. In the absence of final state 
interactions, the phases  $\Dpara$ and $\Dperp$ are both predicted to be 0 or $\pi$ \rad.  
Corrections of order 5\% to these predictions from QCD have been incorporated in more recent 
calculations~\cite{Melikhov:2000yu, Cheng:2001cs}.

The signal decay is flavour specific, with $K^{*0} \to K^+\pi^-$ or $\overline{K}^{*0}\to K^- \pi^+$ 
indicating a $B^0$ or $\overline{B}^0$ decay, respectively.
In the SM, the amplitudes for the decay and its charge-conjugate are equal, but in the presence of physics beyond
the SM (BSM) the loop contributions could be enhanced and introduce
\CP-violating differences between the $B^0$ and $\overline{B}^0$ decay
amplitudes~\cite{london-2004-67, He:1998fk, PhysRevD.71.016007}. 
An analysis of the angular distributions of the decay products gives increased 
sensitivity to BSM physics through differences in the individual amplitudes~\cite{London:ys}.

A further motivation for studying $B^0\to J/\psi K^{*}{^0}$ decays is that the magnitudes and phases of the amplitudes should 
be approximately equal to those in \BsDecay decays~\cite{Gronau2008321}. Both
decay modes are 
dominated by colour-suppressed tree diagrams and have similar branching fractions, 
${\cal{B}}(B^0 \to J/\psi K^{*0}) =  (1.29 \pm 0.14) \times 10
^{-3}$~\cite{BranchingFraction} (S-wave subtracted) and 
\mbox{${\cal{B}}(B_s^0 \to J/\psi \phi) = (1.05 \pm 0.11) \times 10^{-3}$}~\cite{LHCb:bf}.
Any BSM effects observed in $B^0\to J/\psi K^{*}{^0}$ may also be present in $B_s^0 \to J/\psi \phi$, 
where they would modify the time-dependent \CP violation and the \CP-violating 
phase $\phi_s$~\cite{LHCb-PAPER-2013-002}. 

\section{Angular analysis}
\label{sec:Angular}

\begin{figure}
\begin{center}
\includegraphics[scale=0.95, clip=true, trim=30mm 224.5mm 0mm 32.5mm]{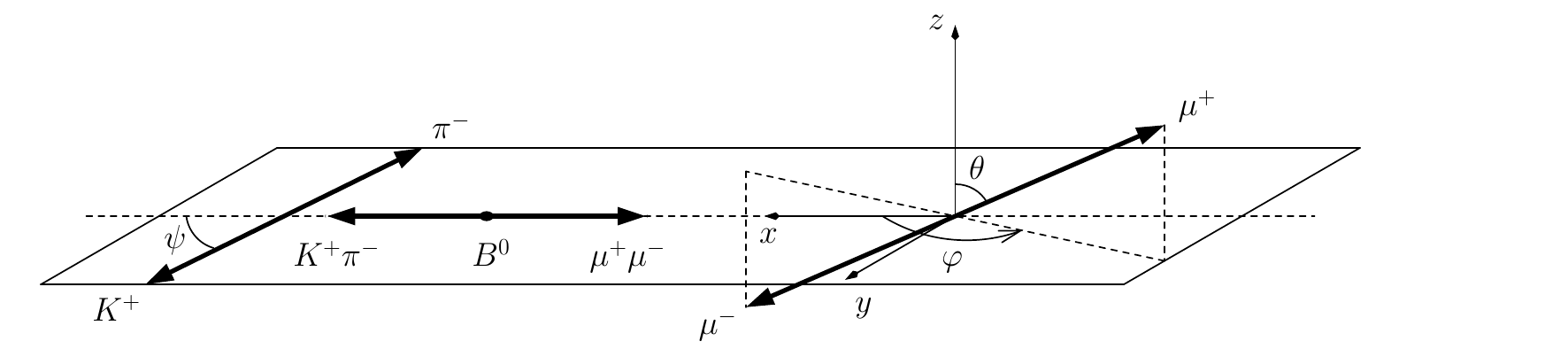}
\caption{\small Definitions of the transversity angles $\theta, \psi,
  \varphi$, as described in the text.}
\label{fig:Trans}
\end{center}
\end{figure}

To measure the individual polarisation amplitudes ($A_0$, $A_{\parallel}$,
$A_{\perp}$, $A_S$) the decay is analysed in terms of 
three angular variables, denoted as $\Omega=\{\cos\theta, \cos\psi, \varphi\}$
in the transversity basis (Fig.~\ref{fig:Trans}).
For a $B^0$ decay, the angle between the $\mu^+$ momentum direction and the
$z$ axis in the $J/\psi$ rest frame is denoted $\theta$ 
and $\varphi$ is the
azimuthal angle of the $\mu^+$ momentum direction in the same frame. $\psi$
is the angle between
the momentum direction of the $K^+$ meson and the negative momentum direction
of the $J/\psi$ meson in the
$K^{*0} \to K^+ \pi^-$ rest frame. For $\overline{B}^0$ decays, the angles are
defined with respect to the $\mu^-$ and the $K^-$ meson.

In this analysis the flavour of the $B$ meson at production is not measured. Therefore,
the observed  $B^0\to J/\psi K^{*}{^0}$ decays arise from both initial $B^0$ or $\overline{B}^0$ mesons as a result of oscillations.
Summing over both contributions, the differential decay rate can be written as~\cite{S-wave,Adeva:2009ny}

\begin{equation}
\begin{split}
\frac{d^4\Gamma(B^0\to J/\psi K^{*}{^0})}{dt\,d\Omega}  \propto e^{-\Gamma_d t}  \sum_{k=1}^{10}  h_k f_k(\Omega)\; ,
\label{eq:diff}
\end{split}
\end{equation}

\noindent where $t$ is the decay time and $\Gamma_d$ is the total decay width of
the $B^0 $ meson; $h_k$ are combinations of the polarisation amplitudes and the $f_k$ are functions of the three 
transversity angles.  These factors can be found in Table~\ref{tab:hf-terms}. 
The $h_k$ combinations are invariant under the phase transformation
$(\delta_\parallel, \delta_\perp, \delta_{\rm S})  \longleftrightarrow (- \delta_\parallel, \pi - \delta_\perp, - \delta_{\rm S})$.
This two-fold ambiguity can be resolved by measuring the phase difference
between the S- and P-wave amplitudes as a function of $m(K^+\pi^-)$ (see Sec.~\ref{sec:results}). The
difference in decay width between the heavy and light eigenstates, 
$\Delta\Gamma_d$, has been neglected.

The differential decay rate for $\overline{B}^0\to J/\psi  \overline{K}^{*0}$
is obtained from Eq.~\ref{eq:diff} by 
defining the angles using the charge conjugate final state particles, and multiplying the interference terms
 $f_4$, $f_6$ and $f_9$ in Table~\ref{tab:hf-terms} by $-1$.
To allow for possible direct \CP violation, the amplitudes are changed from
$A_i$ to $\overline{A}_i$  $(i = 0, \parallel, \perp, {\rm S})$.

\normalsize

\begin{table}[!h]
\begin{center}
\caption{\small Definition of $h_k$ and $f_k$ appearing in
Eq.~\ref{eq:diff}. The $h_k$
factors are invariant under the phase transformation
$(\delta_\parallel, \delta_\perp, \delta_{\rm S}) \longleftrightarrow (- \delta_\parallel, 
\pi - \delta_\perp, - \delta_{\rm S})$~\cite{S-wave,Adeva:2009ny}. The $f_k$ are
functions defined such that their integrals over $\Omega$ are unity.}
\begin{tabular}{c|cc}
$k$             			&     $h_k$
&  $f_k(\Omega) $	\\ \hline \\[-1ex]
1					& $|A_0|^2 $ 							& $\frac{9}{32 \pi} 2 ~\mathrm{cos}^2 \psi (1 - \mathrm{sin}^2 \theta ~\mathrm{cos}^2\varphi$) \\ [6pt]
  2					& $\Aparasq $ 							&	$\frac{9}{32 \pi}~ \mathrm{sin}^2 \psi (1 - \mathrm{sin}^2 \theta~ \mathrm{sin}^2\varphi$) \\ [6pt]
 3					& $\Aperpsq $							& $\frac{9}{32 \pi} ~\mathrm{sin}^2 \psi ~\mathrm{sin}^2 \theta$ \\ [6pt]
4 					&$ |A_{\parallel}| |A_{\perp}| \sin(\Dperp-\Dpara)$	& $-\frac{9}{32 \pi} ~\mathrm{sin}^2 \psi ~\mathrm{sin} 2 \theta ~\mathrm{sin} \varphi$ \\ [6pt]
5					&$ |A_{0}| |A_{\parallel}| \cos(\Dpara)$  		& $\frac{9}{32 \pi \sqrt{2}} ~\mathrm{sin} 2 \psi ~\mathrm{sin}^2 \theta~ \mathrm{sin} 2\varphi$ \\ [6pt]
6					&$ |A_{0}||A_{\perp}| \sin(\Dperp)$  				& $\frac{9}{32 \pi \sqrt{2}}~ \mathrm{sin}~2\psi~ \mathrm{sin}~2 \theta~ \mathrm{cos}~\varphi$ \\ [6pt]
7					&$\Assq $	  							& $\frac{3}{32 \pi} 2 (1 - ~\mathrm{sin}^2 \theta ~\mathrm{cos}^2 \varphi )$\\ [6pt]
8					&$ |A_{\parallel}||A_{\rm S}| \cos(\Dpara - \Ds)$   		&$\frac{3}{32 \pi} \sqrt{6} ~ \mathrm{sin} \psi ~\mathrm{sin}^2 \theta ~\mathrm{sin} 2\varphi$ \\[6pt]
9					&$ |A_{\perp}||A_{\rm S}| \sin(\Dperp - \Ds)$   			& $\frac{3}{32 \pi} \sqrt{6} ~ \mathrm{sin} \psi ~\mathrm{sin}2 \theta ~\mathrm{cos} \varphi$ \\[6pt]
10					&$ |A_{0}||A_{\rm S}| \cos( \Ds)$  							& $\frac{3}{32 \pi} 4 \sqrt{3} ~ \mathrm{cos} \psi (1 - ~\mathrm{sin}^2 \theta ~\mathrm{cos}^2\varphi$)\\[6pt]
\end{tabular}
\label{tab:hf-terms}
\end{center}
\end{table}

\section{LHCb detector}
\label{sec:Detector}

The \lhcb detector~\cite{Alves:2008zz} is a single-arm forward
spectrometer covering the \mbox{pseudorapidity} range $2<\eta <5$, designed 
for the study of particles containing \bquark or \cquark quarks. The 
detector includes a high-precision tracking system consisting of a 
silicon-strip vertex detector surrounding the $pp$ interaction region, 
a large-area silicon-strip detector located upstream of a dipole 
magnet with a bending power of about $4{\rm\,Tm}$, and three stations 
of silicon-strip detectors and straw drift tubes placed 
downstream. The combined tracking system provides a momentum measurement with
relative uncertainty that varies from 0.4\% at 5\gevc to 0.6\% at 100\gevc,
and impact parameter resolution of 20\mum for tracks with high 
transverse momentum ($p_{\mbox{\scriptsize T}}$). Charged hadrons are
identified using two ring-imaging Cherenkov detectors~\cite{Rich}. Photon, electron and
hadron candidates are identified by a calorimeter system consisting of
scintillating-pad and preshower detectors, an electromagnetic
calorimeter and a hadronic calorimeter. Muons are identified by a
system composed of alternating layers of iron and multiwire
proportional chambers.
The trigger consists of a
hardware stage, based on information from the calorimeter and muon
systems, followed by a software stage, which applies a full event
reconstruction.
In the simulation, $pp$ collisions are generated using
\pythia~6.4~\cite{Sjostrand:2006za} with a specific \lhcb
configuration~\cite{gauss}.  Decays of hadronic particles
are described by \evtgen~\cite{Lange:2001uf}, in which final state
radiation is generated using \photos~\cite{photos}. The
interaction of the generated particles with the detector and its
response are implemented using the \geant
toolkit~\cite{Allison:2006ve, *Agostinelli:2002hh} as described in
Ref.~\cite{Clemencic}.


\section{Data samples and candidate selection}
\label{sec:data}

In the following $B^0\to J/\psi K^{*}{^0}$ 
refers to both charge-conjugate decays unless otherwise stated.
The selection of $B^0\to J/\psi K^{*}{^0}$ candidates is based upon the 
decays of the $J/\psi \to \mu^+ \mu^-$ and the $K^{*0} \to K^+ \pi^-$ final states.
Candidates must satisfy the hardware trigger \cite{LHCb-DP-2012-004},
which selects events containing muon candidates that have high transverse momentum
with respect to the beam direction. 
The subsequent software trigger~\cite{LHCb-DP-2012-004} is composed of two
stages. The first stage performs a partial event reconstruction and requires events to 
have two well-identified oppositely-charged muons with invariant mass larger
than $2.7\gevcc$. The second stage of the software trigger performs a full event reconstruction and only retains
events containing a $\mu^+ \mu ^-$ pair that has invariant mass within $120 \,
\mevcc$ of the known $J/\psi$ mass \cite{jr:PDG} 
and forms a vertex that is significantly displaced from the nearest primary
$pp$ interaction vertex (PV).

The $J/\psi$ candidates are formed from two oppositely-charged tracks,
being identified as muons, having $p_{\mbox{\scriptsize T}} > 500 \, \mevc$
and originating from a common vertex. The invariant mass of this pair of muons
must be in the range $3030 - 3150\, \mevcc$.

The $K^{*0}$ candidates are formed from two oppositely-charged tracks, one
identified as a kaon and one as a pion which originate from the same vertex. It is
required that the $K^{*0}$ candidate has  $p_{\mbox{\scriptsize T}} > 2\, \gevc$ and invariant mass in the range $826 - 966 \,\mevcc$.

The $B^0$ candidates are reconstructed from the $J/\psi$ and $K^{*0}$ candidates,
with the invariant mass of the $\mu^+ \mu^-$ pair constrained to the known
$J/\psi$ mass. The resulting $B^0$ candidates are required to have
an invariant mass $m(J/\psi K^+\pi^-)$ in the range $5150 - 5400 \mevcc$.
The decay time of the $B^0$ candidate
is calculated from a vertex and kinematic fit that constrains the $B^0$
candidate to originate from its associated PV~\cite{jr:Hulsbergen:2005pu}. The
$\chi^2$ per degree of freedom of the fit is required to be less
than 5. For events with multiple $B^0$ candidates, the candidate with the
smallest fit $\chi^2$ per degree of freedom is chosen.  Only $B^0$ candidates with
a decay time in the range $0.3 - 14\, \mathrm{ps}$ are retained. The lower
bound on the decay time 
rejects a large fraction of the prompt combinatorial background. 

\begin{figure}
\center{
\includegraphics[width=9.2cm]{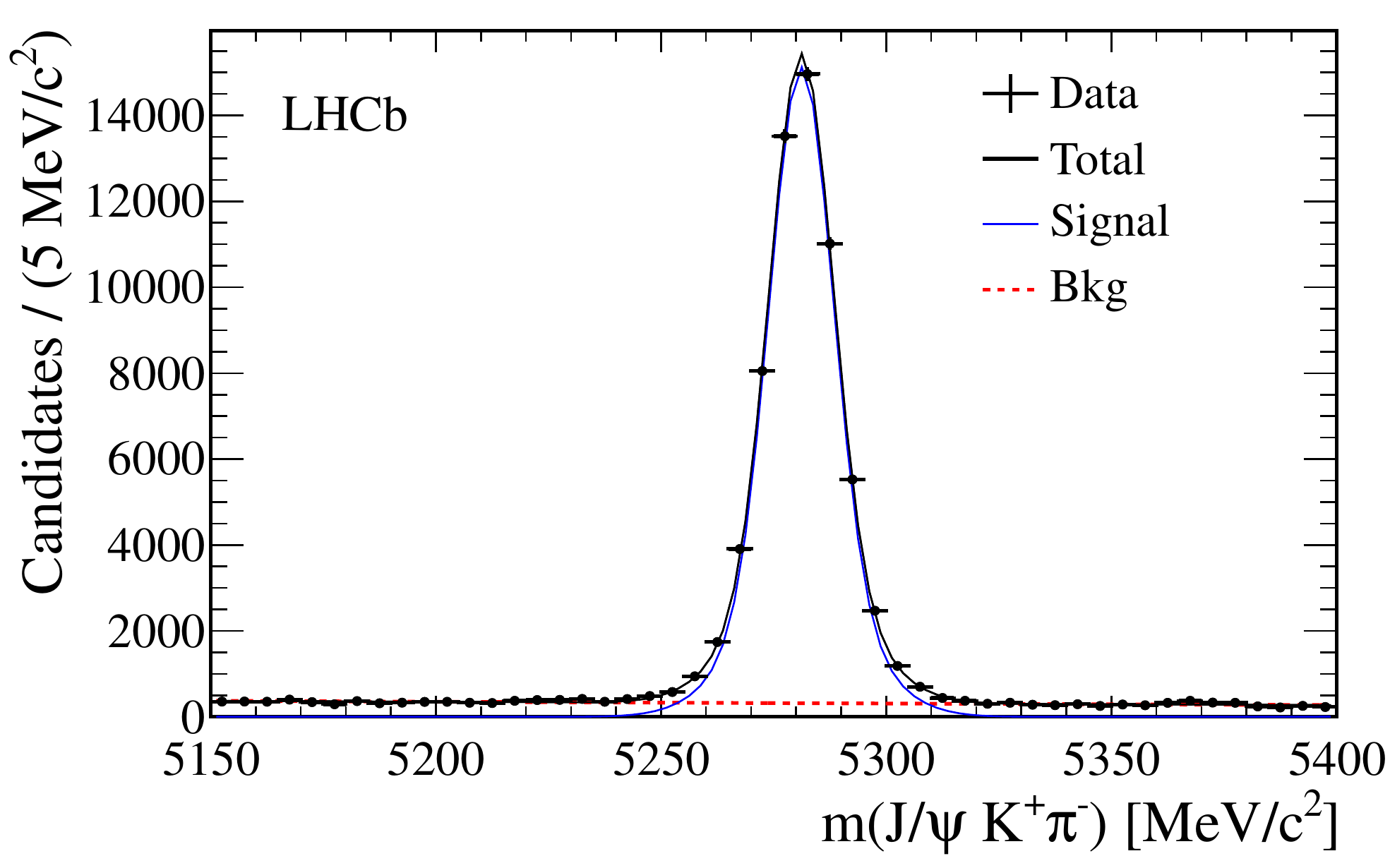}
\caption{\small Invariant mass distribution of the selected $B^0\to J/\psi K^{*}{^0}$ candidates. The curves for the signal 
(solid blue), background (dashed red) and total (solid black) as determined from a fit are shown.}
\label{fig:mass} }
\end{figure}

In the data sample, corresponding to an integrated luminosity of 1.0\invfb,
collected in $pp$ collisions at a centre-of-mass energy of $7$\tev with the
LHCb detector, a total of $77\;282$ candidates are selected. The invariant mass distribution is shown in
Fig.~\ref{fig:mass}. From a fit the number of signal decays is found to be $61\;244 \pm 132$. The
uncertainties on the signal yields quoted here and in Sec.~\ref{sec:results} come from propagating 
the uncertainty on the signal fraction evaluated by the fit. 

\section{Maximum likelihood fit}
\label{sec:fit}

The parameters used in this analysis are $\Aparasq,\Aperpsq, \Fs, \Dpara, \Dperp$ and $\Ds$, 
where we introduce the parameter $\Fs =  \Assq / (1 +  \Assq)$  to
denote the fractional S-wave component.
The parameter $\Azerosq$ is determined by the constraint  $\Azerosq + \Aparasq
+\Aperpsq = 1$.
The best fit values of these parameters are determined with an
unbinned maximum log-likelihood fit to the
decay time and angular distributions of the selected $B^0$
candidates. In order to subtract the background component, 
each event is given a signal weight, $W_i$, using the $sPlot$
\cite{Pivk:2004ty} method with $m(J/\psi K^+\pi^-)$ as the discriminating
variable. The invariant mass distribution of the
signal is modelled as the sum of two Gaussian functions with a common mean. 
The mean and widths of both Gaussian functions, as well as the fraction of the first
Gaussian are parameters determined by the fit. The effective resolution of the mass
peak is determined to be $9.3 \pm 0.8 \mevcc$. The invariant mass
distribution of the background is described by an exponential function.
The signal fraction in a $\pm30\,\mevc$ window around
the known $B^0$ mass~\cite{jr:PDG} is approximately $93\%$.

A maximum likelihood fit is then performed with each candidate weighted by $W_i$. 
The fit uses a signal-only probability density function (PDF)  which is denoted $\mathcal{S}$. 
It is a function of the decay time $t$ and angles $\Omega$, and is obtained from Eq.~\ref{eq:diff}.
The exponential decay time function is convolved with a Gaussian function 
to take into account the decay time resolution of 45\fs~\cite{LHCb-PAPER-2013-002}. 
The effect of the time and angular  resolution on this analysis has been studied and found to be negligible \cite{Adeva:2009ny}.

The fit minimises the negative log likelihood summed over the selected candidates

\begin{equation}
- \ln \,\mathcal{L} = - \alpha \sum_{i} W_i \, \ln \, \mathcal{S}_i(t_i, \Omega_i) \;,
\end{equation}

\noindent where  $\alpha = \sum_{i} W_i /
\sum_{i} W_i^2$ is a normalisation factor accounting for the effect of the weights in the determination
of the uncertainties \cite{Aaij:2011qx}. 

The selection applied to the data is almost unbiased with respect to the decay
time. 
The measurements of amplitudes and phases are insensitive to the decay time 
acceptance since 
$\Delta \Gamma_d \sim 0$ and  the time dependence of the PDF factorises out from the angular part.
Nevertheless, the small deviation of the decay time acceptance from uniformity
is determined from data using decay time unbiased triggers as a reference, 
and is included in the fitting procedure. 

\begin{figure}
\begin{center}
\subfloat{
\includegraphics[width=8.2cm]{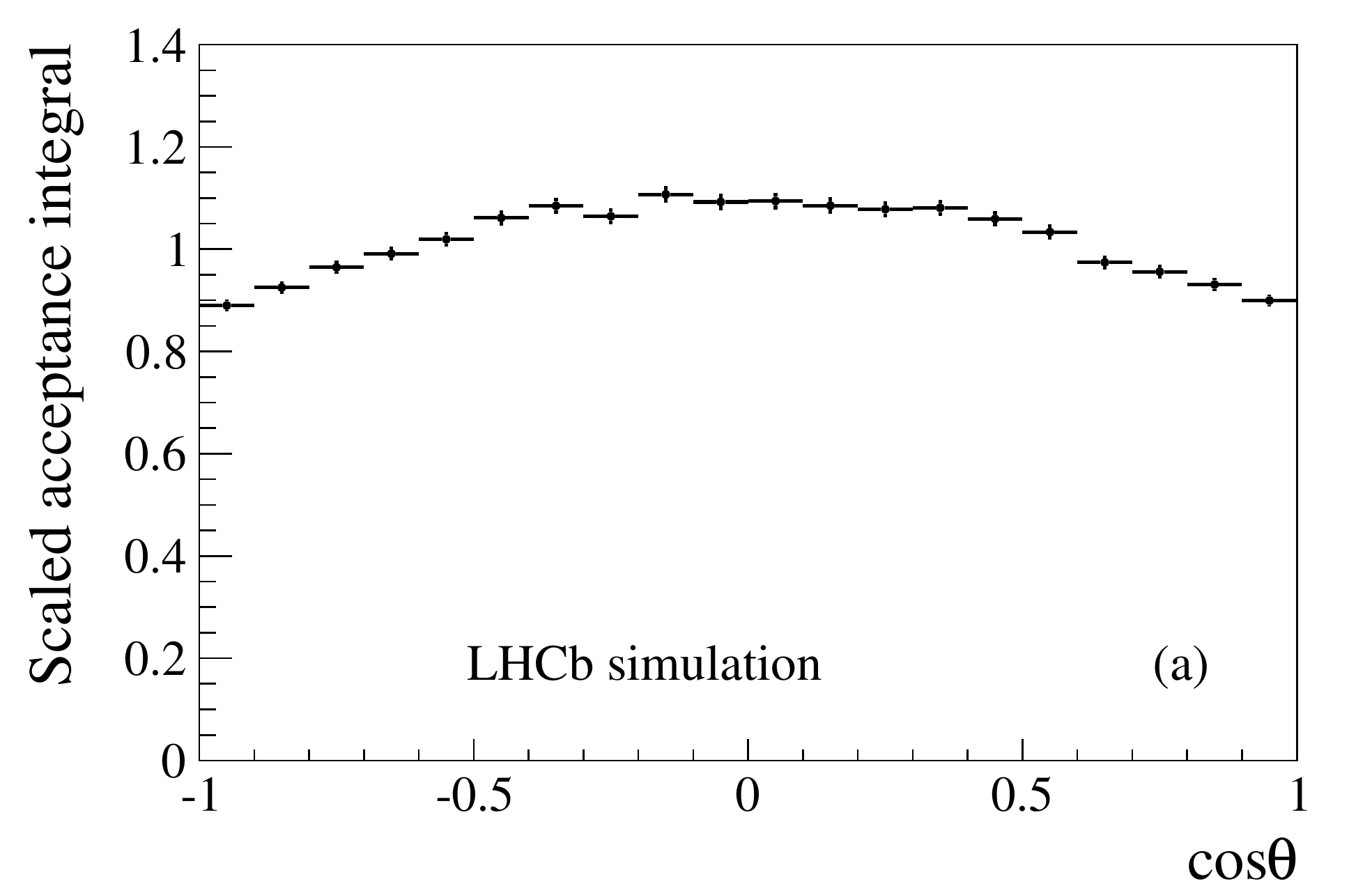}}
\subfloat{
\includegraphics[width=8.2cm]{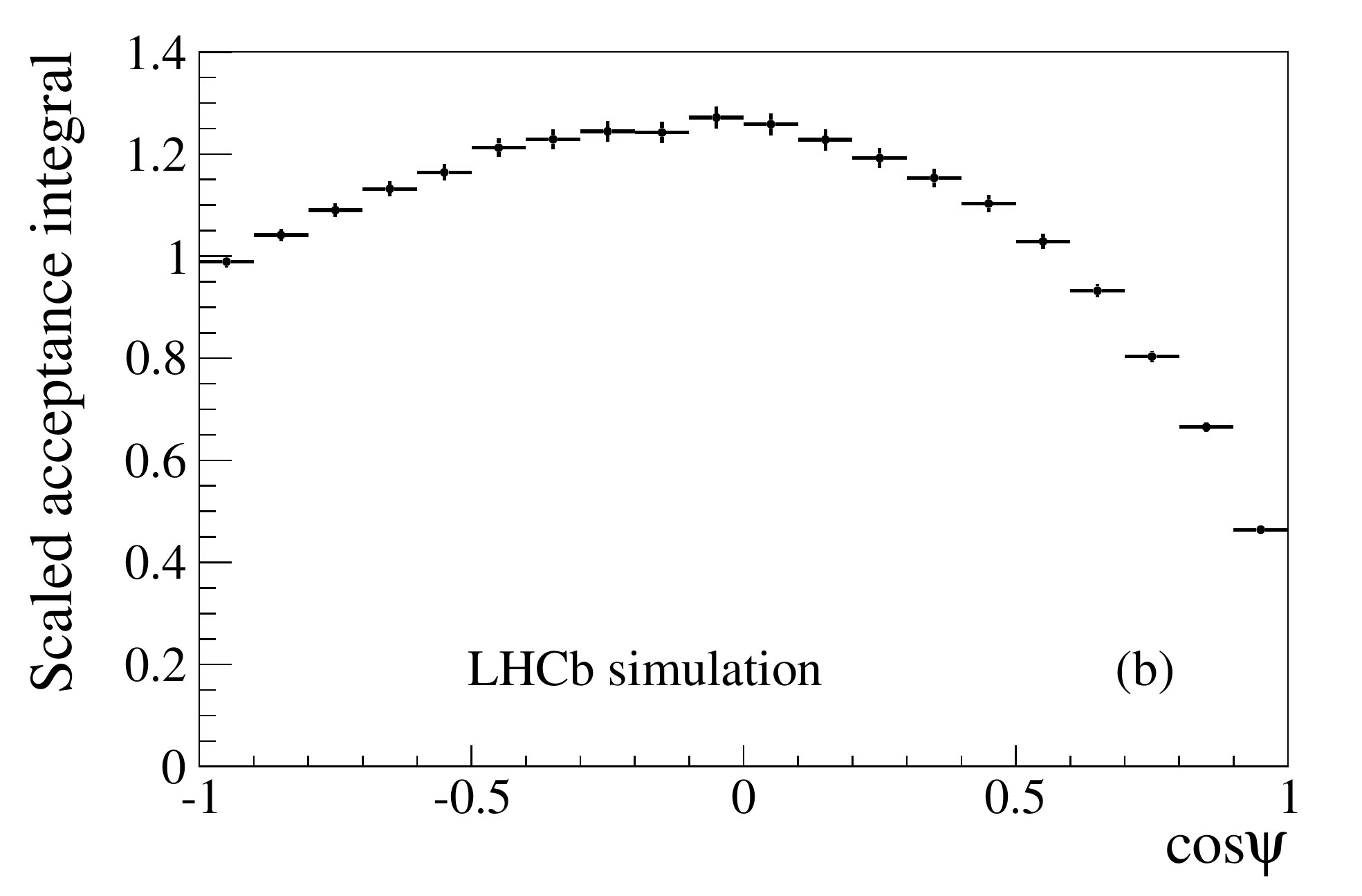}}\\
\subfloat{
\includegraphics[width=8.2cm]{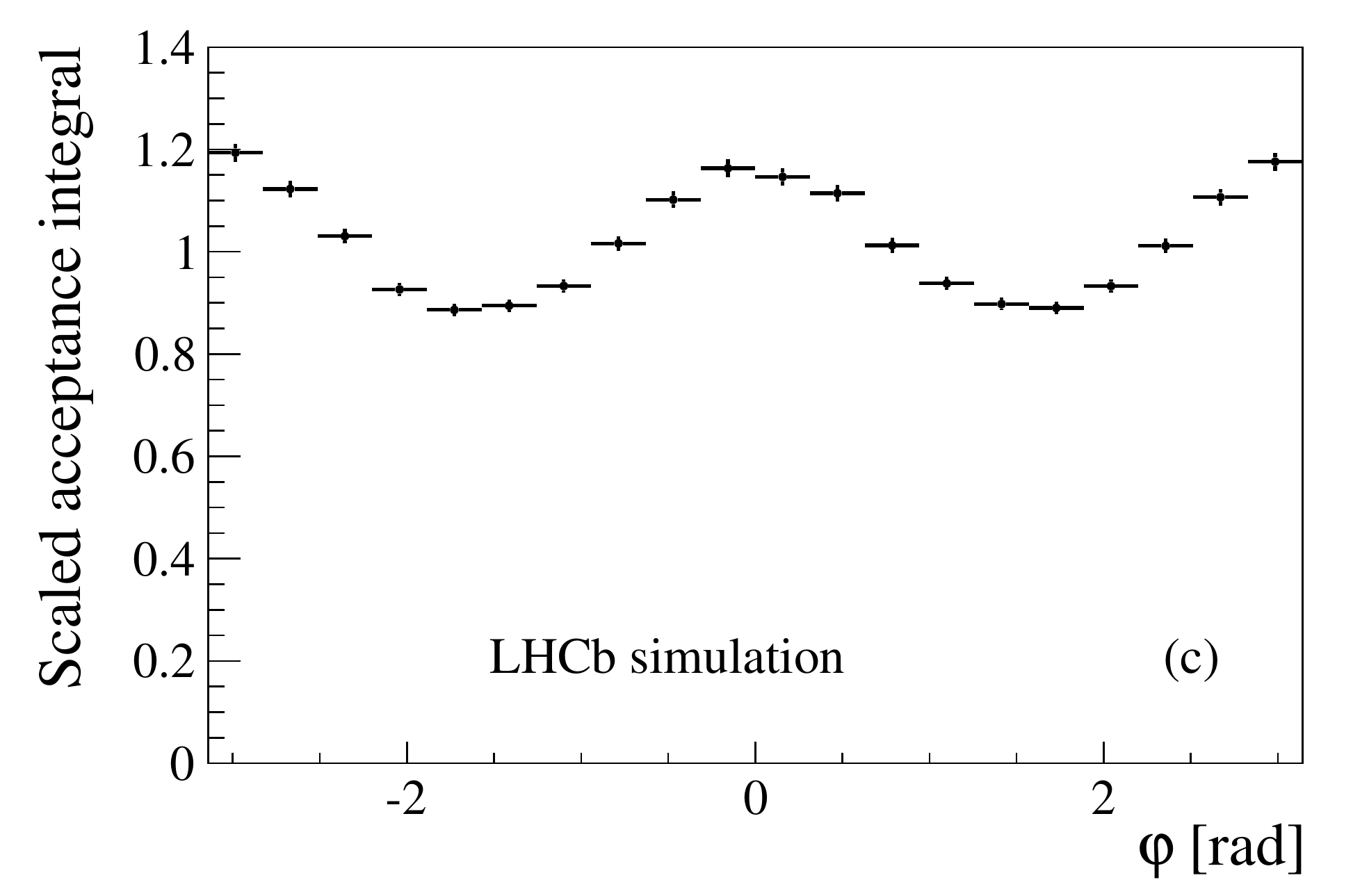}}
\end{center}
\caption{\small Angular acceptance $A(\Omega)$ as a function of each decay
  angle, integrated over the other two angles for (a) $\cos\theta$, (b)
  $\cos\psi$ and (c) $\varphi$. The projections are normalised such that their 
  average value over the histogram range is unity.}
\label{fig:acceptance}
\end{figure}

The acceptance as a function of the decay angles is not uniform because of the forward geometry of the detector 
and the momentum selection requirements applied to the final state particles. 
A three-dimensional acceptance function, $A(\Omega)$, is determined using simulated events subject 
to the same selection criteria as the data, and is included in the fit.
Figure~\ref{fig:acceptance} shows the acceptance as a function of each decay
angle, integrated over the two other angles.
The variation in acceptance is asymmetric for
$\cos\psi$, due to the selection requirements 
on the $\pi^-$ and the $K^{*0}$ mesons.

\begin{figure}[t]
\begin{center}
\includegraphics[width=8cm]{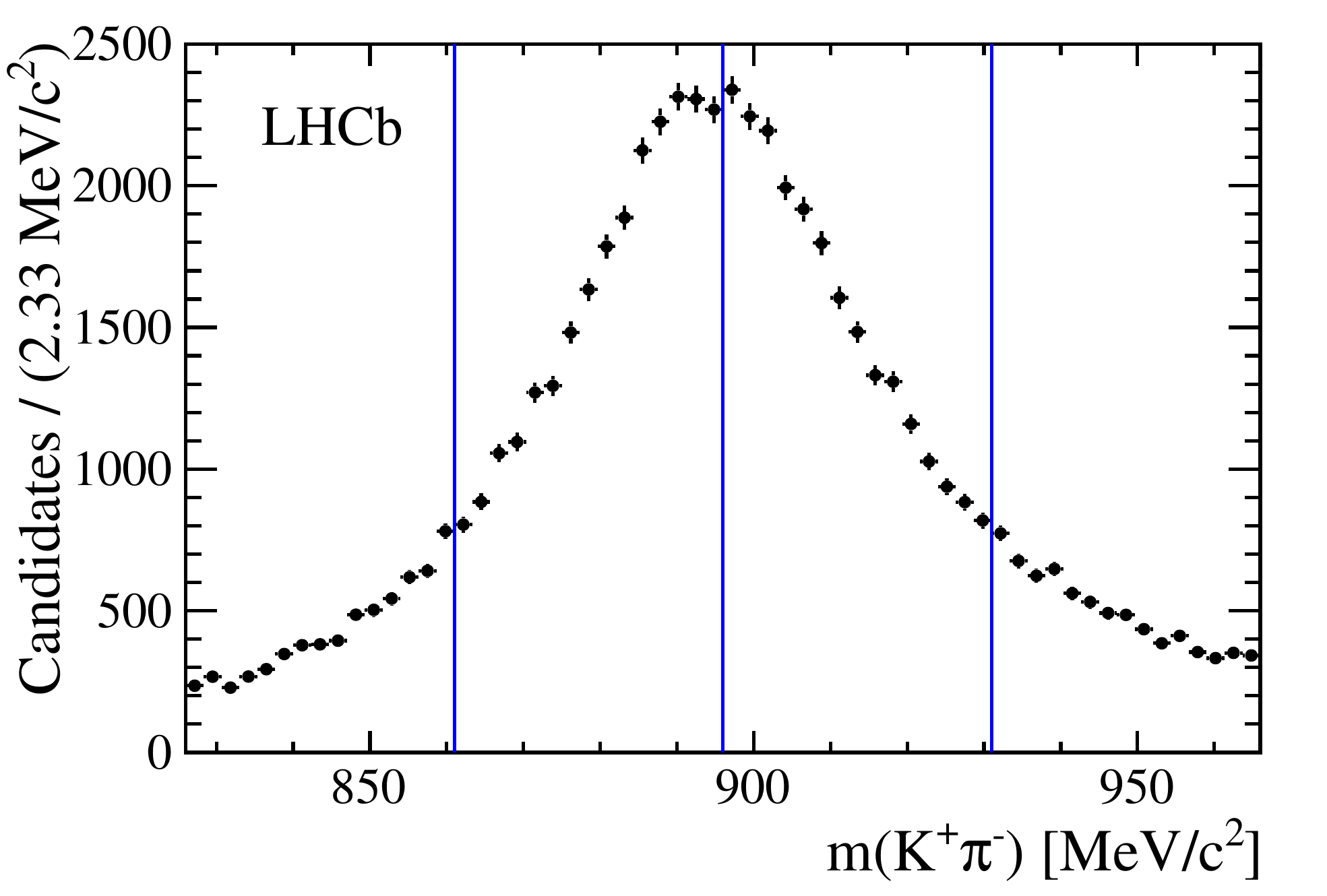}
\caption{\small Background subtracted distribution of the $m(K^+\pi^-)$ invariant mass. The four bins used to 
resolve the ambiguity in the strong phases are shown.}
\label{fig:KpiBins}
\end{center}
\end{figure}

The phase of the P-wave amplitude increases rapidly as a
function of the $K^+\pi^-$ invariant mass, whereas the S-wave phase increases
relatively slowly~\cite{Aubert:2005uq}. As a result the phase difference
between the S- and P-wave amplitudes falls with increasing $K^+\pi^-$ invariant
mass. A fit which determines the phase difference in bins of $m(K^+\pi^-)$ can therefore be used to select the
physical solution and hence resolve the ambiguity described in Sec.~\ref{sec:Angular}.  This method has previously been
used to measure the sign of $\Delta \Gamma_s$ in the $B_s^0$ system
~\cite{Collaboration:2012uq}.
In the analysis the data are divided into four bins of $m(K^+\pi^-)$, shown in 
Fig.~\ref{fig:KpiBins} and defined in Table~\ref{tab:Kpibins}.
A simultaneous fit to all four bins is performed in which the P-wave parameters are
common, but $\Fs$ and $\Ds$ are independent parameters in each bin.
Consistent results are obtained with the use of two or six bins.

To correct for the variation of the S-wave relative to
the P-wave over the $m(K^+\pi^-)$ range of each bin, a correction factor is introduced in each of the three
interference terms $f_8$, $f_9$ and $f_{10}$ in Eq.~\ref{eq:diff}. The S-wave
lineshape is assumed to be uniform across the $m(K^+\pi^-)$ range
and the P-wave shape is described by a relativistic Breit-Wigner function. 
The correction factor is calculated by integrating the product $p s^*$ 

\begin{equation}
\int^{m^{H}_{K^+\pi^-}}_{m^L_{K^+\pi^-}} p  s^* ~ dm(K^+\pi^-) = C_{\rm SP} e^{-i\theta_{\rm SP}} \; ,
\end{equation}

\noindent 
where $p$ and $s$ are the P- and S-wave lineshapes normalised to unity in the
range of integration, * is the complex conjugation operator, 
$m^L_{K^+\pi^-}$ and $m^H_{K^+\pi^-}$ denote the boundaries of the
$m(K^+\pi^-)$ bin, $C_{\rm SP}$ is the correction factor and $\theta_{\rm SP}$ is
absorbed in the measurements of $\Ds - \Dzero$.  
The $C_{\rm SP}$ factors tend to unity (i.e. no correction) as the bin width tends to zero.
The $C_{\rm SP}$ factors calculated for this analysis are given in Table~\ref{tab:Kpibins}. 
The factors are close to unity, and hence the analysis
is largely insensitive to this correction. 

\begin{table}
\caption{\small Bins of $m(K^+\pi^-)$ and the corresponding $C_{\rm SP}$ correction factor for 
the S-wave interference terms, assuming a uniform distribution for the non-resonant $K^+\pi^-$ 
contribution and a relativistic Breit-Wigner shape for decays via the $K^{*0}$ resonance.}
\begin{center}
\begin{tabular}{c|c}
$m(K^+\pi^-)$ [$\mevcc$] & $C_{\rm SP}$ \\
\hline
826 -- 861 & 0.984 \\
861 -- 896 & 0.946 \\
896 -- 931 & 0.948 \\
931 -- 966 & 0.985 \\
\end{tabular}
\end{center}
\label{tab:Kpibins}
\end{table}


\section{Systematic uncertainties}
\label{sec:systematics}

To estimate the systematic uncertainties arising from the choice of the model
for the $B^0$ invariant mass, the signal mass PDF 
is changed from a double Gaussian function
to either a single Gaussian or a Crystal Ball function. The largest differences observed 
in the fitted values of the parameters
are assigned as systematic uncertainties.

To account for uncertainties in the treatment of the combinatorial background,
an alternative fit to the data is performed without using signal weights. 
An explicit background model, $\mathcal{B}$, is constructed, with the time distribution being 
described by two exponential functions,  
and  the angular distribution by a three-dimensional histogram derived from the sidebands of 
the $B^0$ invariant mass distribution. 
A fit is then made to the unweighted data sample with the sum of $\mathcal{S}$ and $\mathcal{B}$.
The results of this fit are consistent with those from the fit using signal weights and the small 
differences are included as systematic uncertainties.
 
\begin{table}[!h]
\caption{\small Systematic uncertainties as described in the text. The contribution from 
omitting the $C_{\rm SP}$ factors is negligible for the P-wave parameters. The total systematic
uncertainty is the sum in quadrature of the individual contributions.}
\begin{center}
\subfloat[P-wave parameters]{
\begin{tabular}{c|c|c|c|c}
Source & $\Aparasq$ & $\Aperpsq$ & $\Dpara$[rad] & $\Dperp$[rad] \\
\hline
Mass model	                      & 0.000 & 0.001 & 0.00 & 0.00  \\
Background treatment	              & 0.002 & 0.001 & 0.00 & 0.00  \\
Misreconstructed background          & 0.002 & 0.000 & 0.00 & 0.01  \\
Angular acceptance                    & 0.009 & 0.007 & 0.03 & 0.01  \\
Statistical uncertainty on acceptance & 0.001 & 0.001 & 0.01 & 0.01  \\
Other resonances                    & 0.005 & 0.004 & 0.00 & 0.01  \\
\hline
Total systematic uncertainty          & 0.011 & 0.008 & 0.03 & 0.02  \\
\hline 
Statistical uncertainty               & 0.004 & 0.004 & 0.02 & 0.02  \\
\end{tabular}
}\\
\vspace{0.2cm}
\subfloat[S-wave parameters of bins (1) and (2)]{
\begin{tabular}{c|c|c|c|c}
Source & $\Fs^{(1)} $ & $\Ds^{(1)}$[rad]  & $\Fs^{(2)}$   & $\Ds^{(2)}$[rad] \\
\hline
Mass model	                      & 0.005 & 0.01 & 0.001 & 0.01  \\
Background treatment	              & 0.003 & 0.04 & 0.001 & 0.01  \\
Misreconstructed background          & 0.006 & 0.01 & 0.002 & 0.00  \\
Angular acceptance                    & 0.007 & 0.01 & 0.004 & 0.05  \\
Statistical uncertainty on acceptance & 0.003 & 0.04 & 0.002 & 0.03  \\
$C_{\rm SP}$ factors                  & 0.003 & 0.00 & 0.005 & 0.01  \\
Other resonances                      & 0.016 & 0.06 & 0.002 & 0.02  \\
\hline
Total systematic uncertainty          & 0.020 & 0.08 & 0.007 & 0.06  \\
\hline 
Statistical uncertainty               & 0.007 & 0.10 & 0.004 & 0.06  \\
\end{tabular}
}\\
\vspace{0.2cm}
\subfloat[S-wave parameters of bins (3) and (4)]{
\begin{tabular}{c|c|c|c|c}
Source & $\Fs^{(3)} $ & $\Ds^{(3)}$[rad]  & $\Fs^{(4)}$   & $\Ds^{(4)}$[rad] \\
\hline
Mass model	                      & 0.003 & 0.01 & 0.004 & 0.01  \\
Background treatment	              & 0.001 & 0.01 & 0.003 & 0.02  \\
Misreconstructed background          & 0.003 & 0.01 & 0.004 & 0.01  \\
Angular acceptance                    & 0.000 & 0.08 & 0.003 & 0.05  \\
Statistical uncertainty on acceptance & 0.002 & 0.03 & 0.003 & 0.04  \\
$C_{\rm SP}$ factors                  & 0.005 & 0.00 & 0.002 & 0.00  \\
Other resonances                      & 0.006 & 0.02 & 0.000 & 0.08  \\
\hline
Total systematic uncertainty          & 0.009 & 0.09 & 0.008 & 0.11  \\
\hline 
Statistical uncertainty               & 0.006 & 0.03 & 0.014 & 0.03  \\
\end{tabular}
}
\end{center}
\label{systematics}
\end{table}

A very small contribution from the decay $B_s^0 \to J/\psi \overline{K}^{*0}$~\cite{LHCb:2012wx}
in the high-mass sideband of the $B^0$ invariant mass distribution of
Fig.~\ref{fig:mass} has a negligible
effect on the fit results. The only significant background that peaks in the
$B^0$ mass region arises from candidates where one or more of the tracks are 
misreconstructed, in most of the cases the pion track.
From simulation studies we find that this corresponds to 3.5\%
of the signal yield and has a similar $B^0$ mass distribution
to the signal but a significantly different angular
distribution. The yield and shape of the background are taken from simulated
events, and are used to explicitly model this background in the data fit. The effect on the fit results 
is taken as a systematic uncertainty. Other background contributions are found to be insignificant.

The angular acceptance function is determined from simulated events, and a systematic uncertainty is 
included to take into account the limited size of the simulated event sample.
An observed difference in the kinematic distributions of the final state particles
between data and simulation is largely attributed to the S-wave component, which is not included in the simulation.
To account for the S-wave, the simulated events are reweighted to match 
the signal distributions expected from the best estimate of the physics
parameters from data (including the S-wave).  After this procedure, small differences remain in the pion and kaon
momentum distributions. The simulated events are further reweighted to remove
these differences, and the change in the fit results is taken as the systematic uncertainty due 
to the modelling of the acceptance.  

The $C_{\rm SP}$ factors do not affect the P-wave amplitudes and only have a small
effect on the S-wave amplitudes. The fit is performed with each $C_{\rm SP}$ factor set to unity, and
the differences in the S-wave parameters are taken as a systematic uncertainty.

This analysis assumes only P- and S-wave contributions to the $K^+\pi^-$ system, but makes no assumption about
the $m(K^+\pi^-)$ mass model itself (except in the determination of the
$C_{\rm SP}$ factors). The S-wave fractions reported in Table~\ref{tab:simFit} correspond  
to a shape that does not exhibit an approximately linear S-wave (as might be na\"ively expected). 
A separate study of the $m(K^+\pi^-)$ mass spectrum and angular distribution has been performed over a 
wider $m(K^+\pi^-)$ mass range. 
This study indicates that there may be contributions from additional
resonances, e.g. $\kappa(800)$, $K^{*}(1410)$, $K^{*}_{2}(1430)$ and
$K^{*}(1680)$ states. Of particular interest is the $K^*_2(1430)$ contribution, which is a D-wave state and can interfere with the P-wave. 
Using simulated pseudo experiments such interferences are observed to change
the shape of the observed $m(K^+\pi^-)$ spectrum from that corresponding to a
simple linear S-wave, and that by ignoring such possible additional resonances the P- and S-wave parameters may be biased. 
These biases are estimated using simulated experiments containing these additional resonances 
and they are assigned as systematic uncertainties. The systematic uncertainties are summarised in Table \ref{systematics}.


\section{Results}
\label{sec:results}

\begin{figure}[!h]
\begin{center}
\subfloat{
\includegraphics[width=8cm]{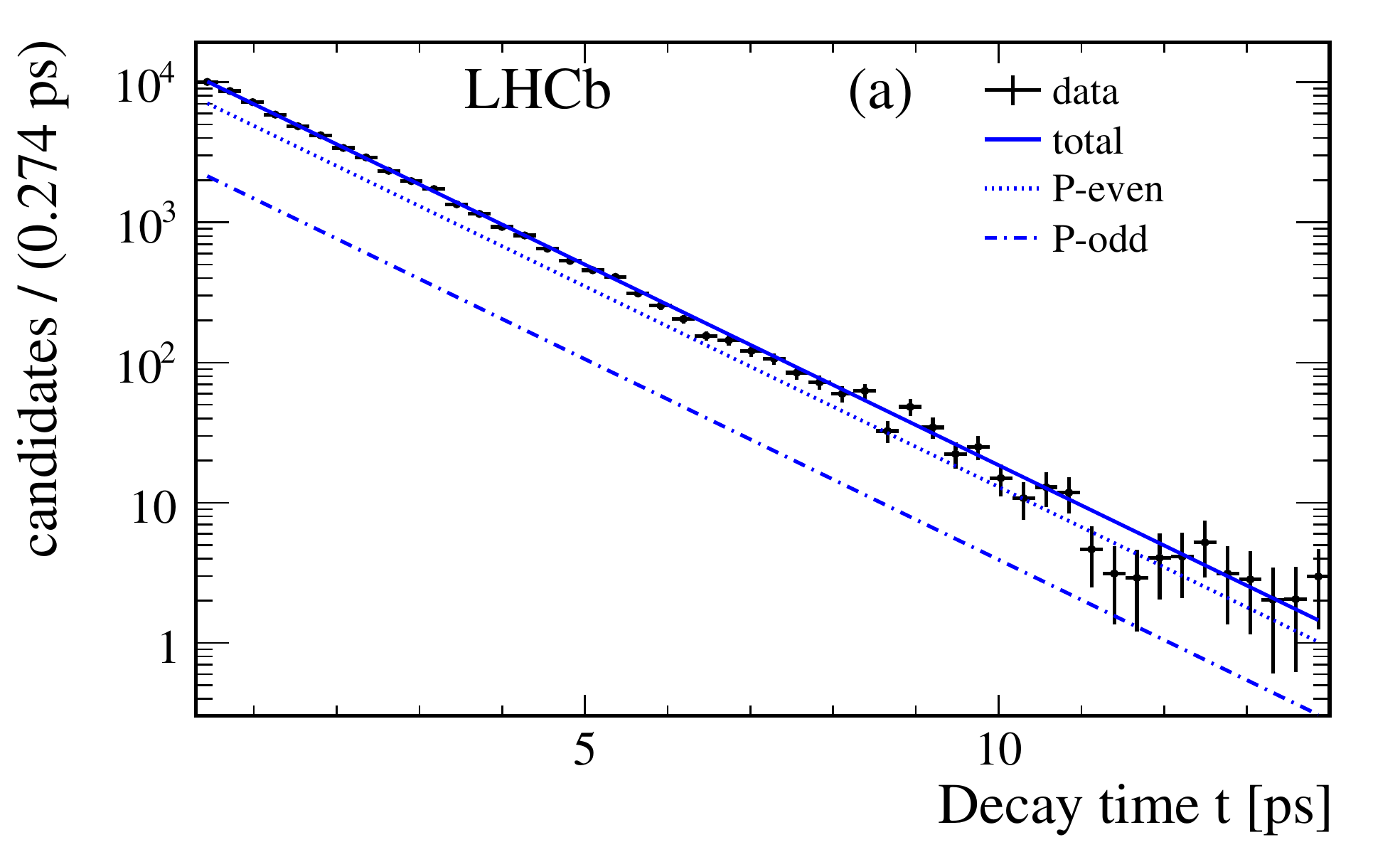}
}
\subfloat{
\includegraphics[width=8cm]{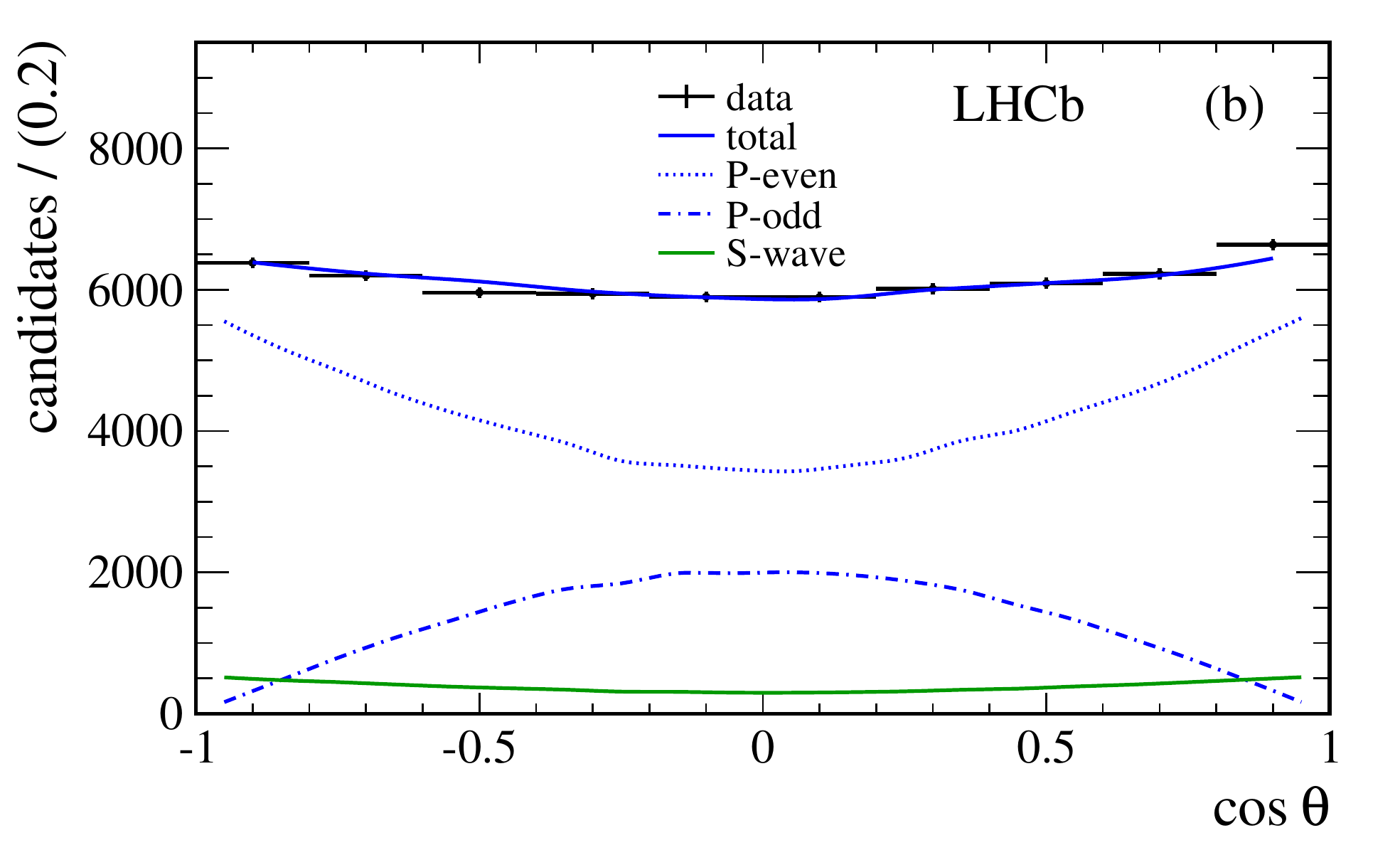}
}

\subfloat{
\includegraphics[width=8cm]{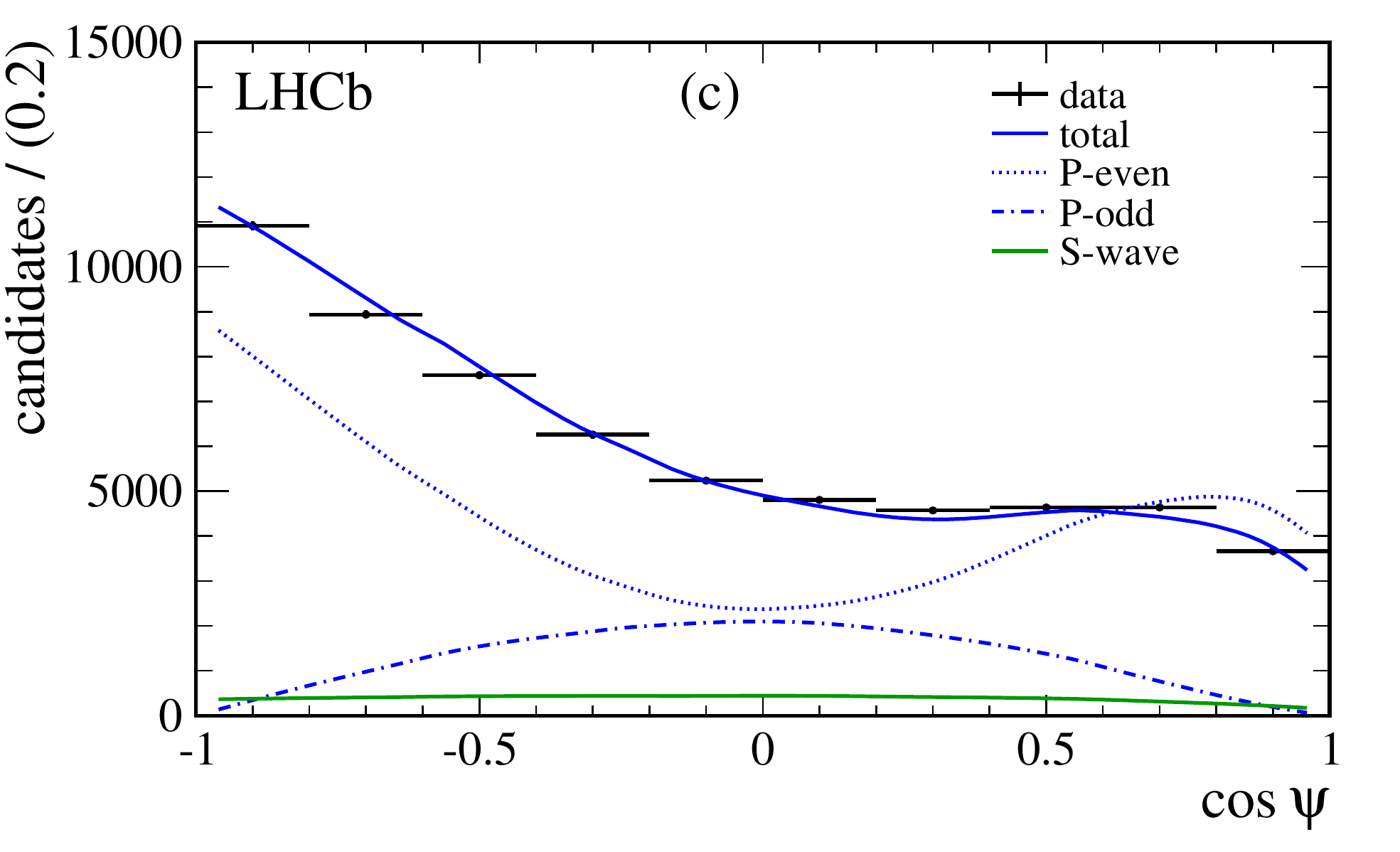}
}
\subfloat{
\includegraphics[width=8cm]{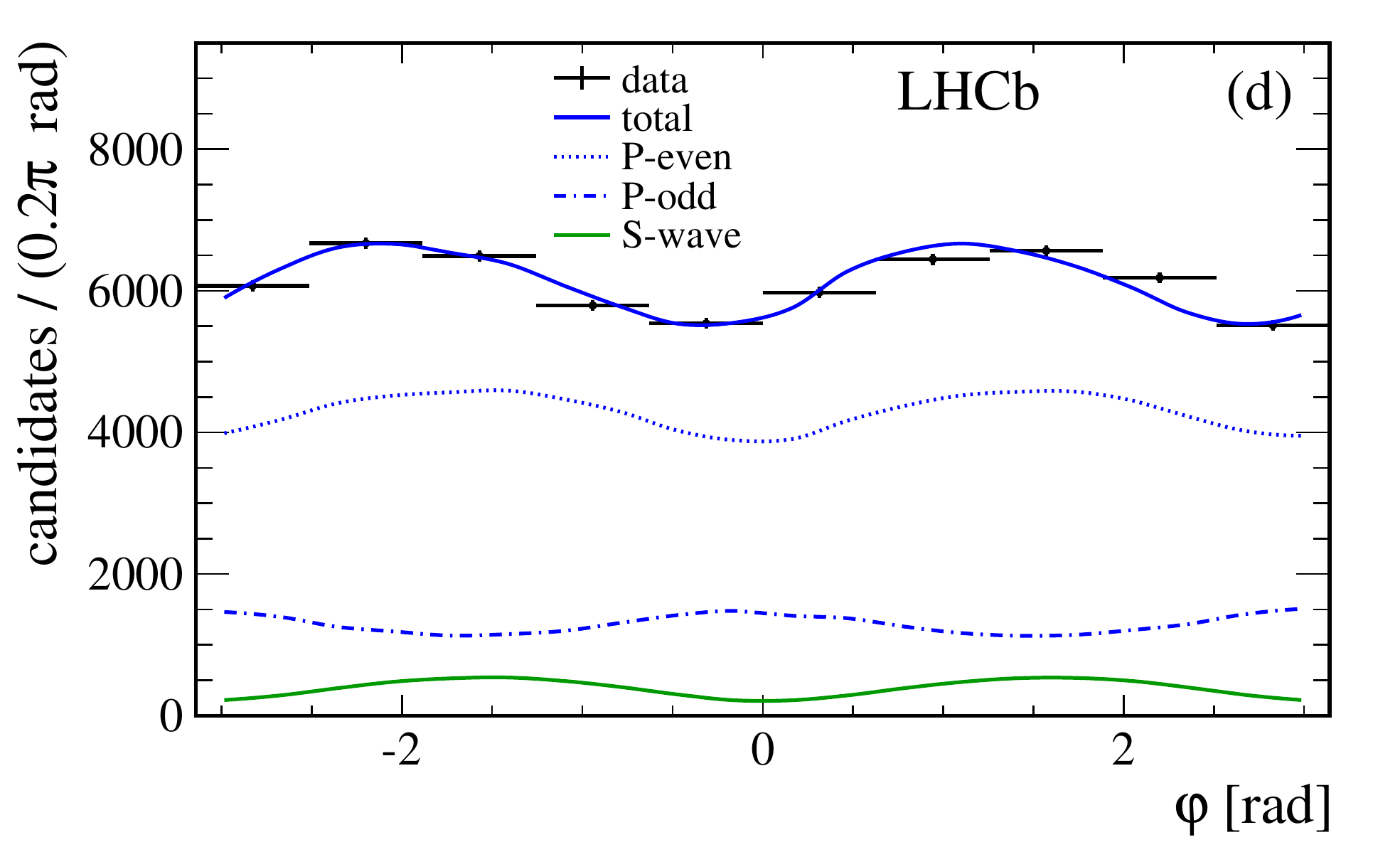}
}
\end{center}
\caption{\small Projections of (a) the decay time and the transversity angles
(b) $\cos\theta$, (c) $\cos\psi$ and (d) $\varphi$ from the fit to the data
(points with statistical error bars). The
different curves show the P-wave parity-even (dotted blue) and parity-odd (dashed blue)
components, the pure S-wave (green) contributions without interference,
as well as the total signal component (solid blue).}
\label{fig:angles}
\end{figure}

\begin{figure}[!h]
\begin{center}
\subfloat{
\includegraphics[width=7.8cm]{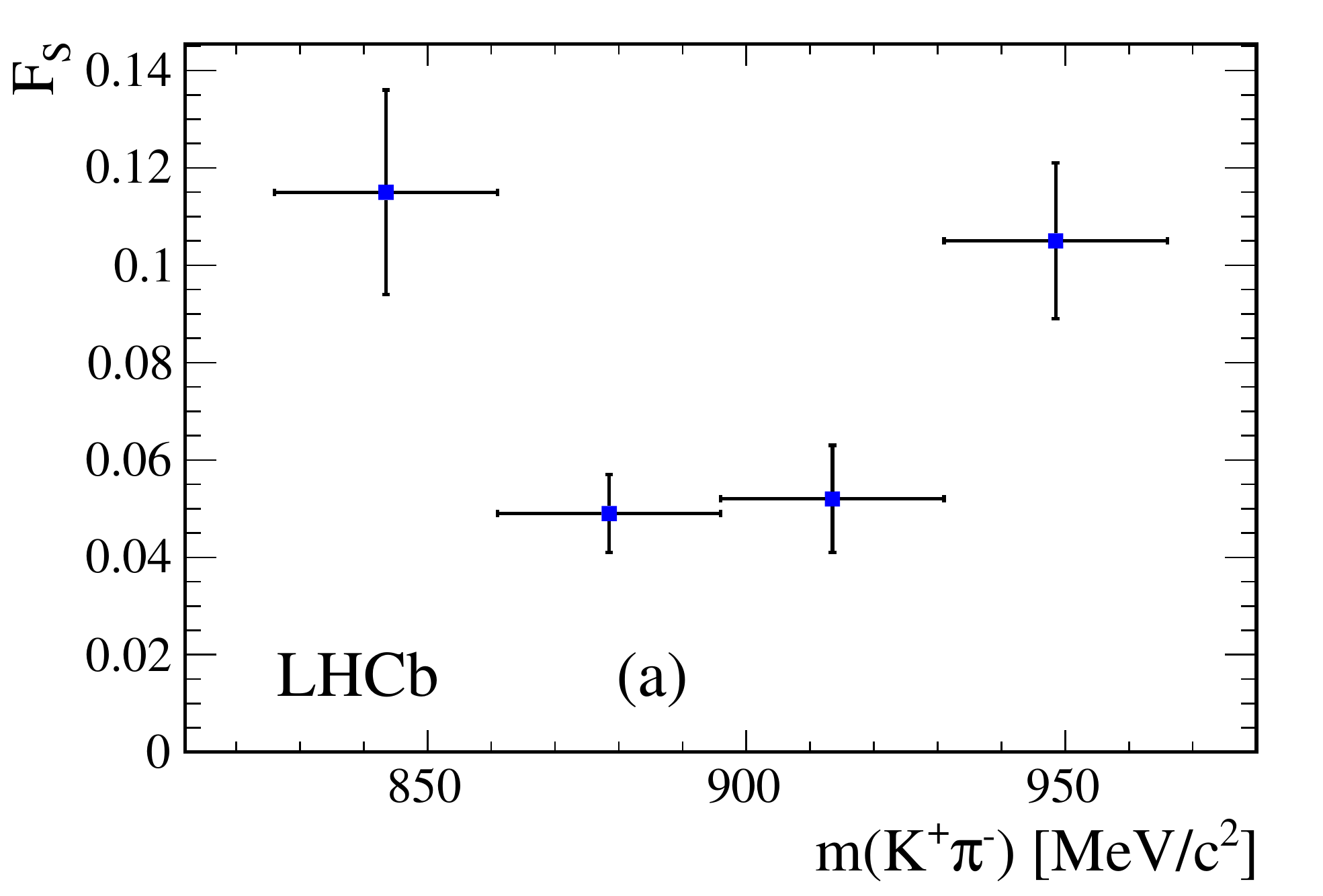}
}
\subfloat{
\includegraphics[width=7.8cm]{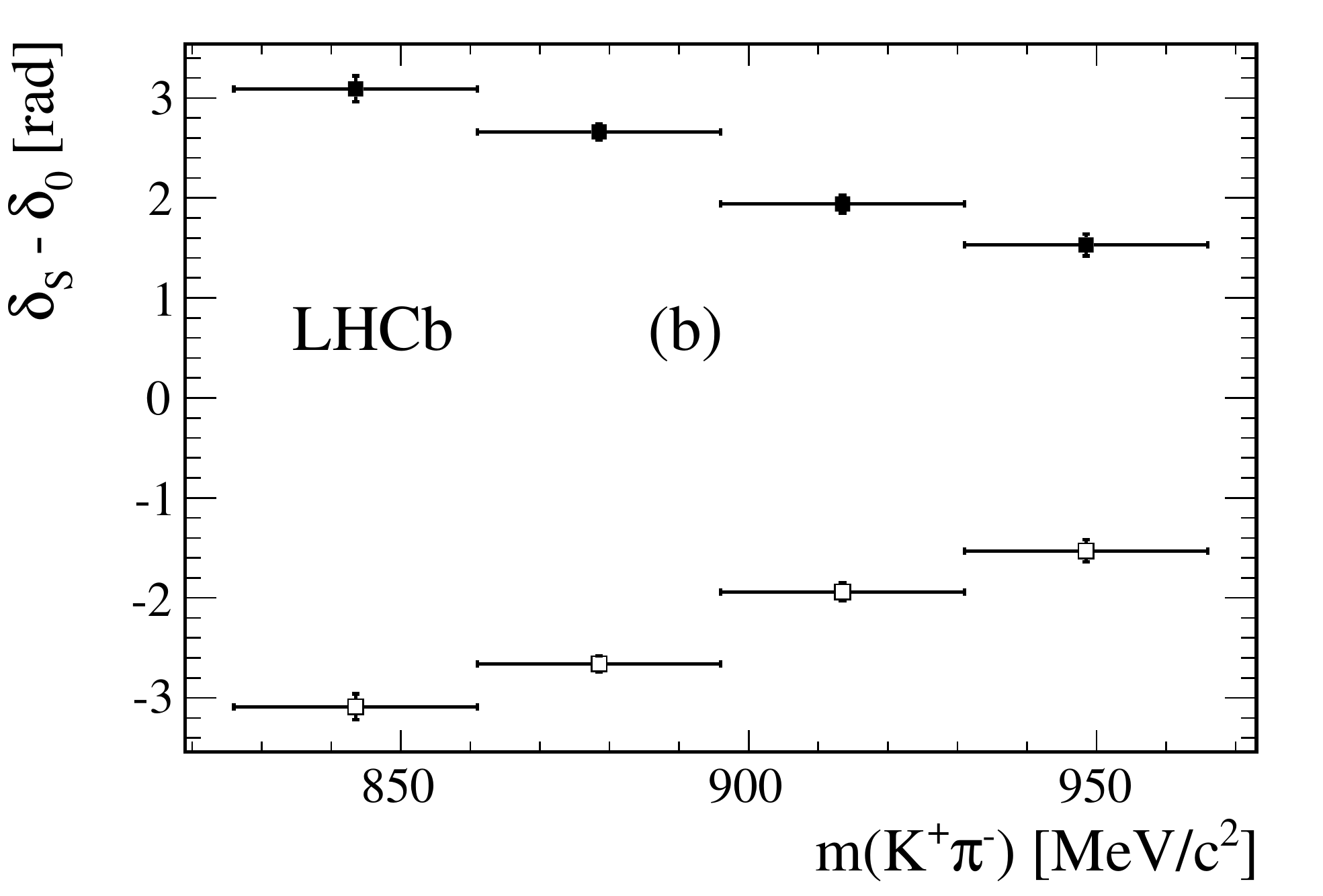}
}
\caption{\small Variation of (a) $F_{\rm S}$ and (b) $\delta_{\rm S} -
  \delta_{\rm 0}$ in the simultaneous fit in four bins of the
  $K^+\pi^-$ mass. There are two solutions of the relative phase, the falling
  trend (solid points) being the physical one.}
\label{fig:simFit}
\end{center}
\end{figure}

The values of the P-wave parameters obtained from the fit
to the combined $B^0 \to J/\psi K^{*0}$ and $\overline B^0 \to J/\psi \overline K^{*0}$ samples,
assuming no direct \CP violation, are
shown in Table~\ref{FinalRes} with their statistical and systematic
uncertainties.
The projections of the
decay time and the transversity angles are shown in Fig.~\ref{fig:angles}.
Although we have included the decay time distribution in the fit, we do not
report a lifetime measurement here, which will instead be
included in a forthcoming publication.
Figure~\ref{fig:simFit} shows the values for $\Fs$ and $\Ds-\Dzero$ as a
function of the $K^+\pi^-$
mass. The phase $\Dzero=0$ is inserted explicitly to emphasise that this is
the phase difference
between the S- and P-waves.
The error bars include both the statistical and systematic uncertainties.
The solid points of Fig.~\ref{fig:simFit}(b) correspond to the physical
solution with a decreasing phase difference. Table~\ref{tab:simFit} presents
the
values of $\Fs$ and $\Ds - \delta_0$ for the physical solution. The
correlation matrix for the P- and S-wave parameters is shown in Table~\ref{tab:corr}.
Integrating the S-wave fraction over all four $m(K^+\pi^-)$ bins gives an
average value of \mbox{$F_S=(6.4 \pm                                                             
0.3 \pm 1.0)\%$} in the full window of $\pm$70\mevcc around the known
$K^{*}{^0}$ mass~\cite{jr:PDG}. The BaBar collaboration
\cite{Collaboration:2007kx} measured an S-wave component of \mbox{$(7.3 \pm
  1.8)\%$}
in \mbox{$B^0 \to J/\psi K^+\pi^-$} in a $K^+\pi^-$ mass range from 0.8 to 1.0 \gevcc.

\begin{table}[!h]
\caption{\small Results for \mbox{$B^0\to J/\psi K^{*0}$} candidates. The uncertainties are statistical and systematic, respectively.}
\begin{center}
\begin{tabular}{c|c}
           Parameter &    Value    \\
\hline
$\Aparasq$ 			&         0.227 $\pm$   0.004 $\pm$ 0.011     \\
$\Aperpsq$ 			&         0.201 $\pm$    0.004        $\pm$ 0.008       \\
$\Dpara$ [rad] 			&         $-$2.94 $\pm$     0.02 $\pm$ 0.03\\
$\Dperp$ [rad]			&        $\phantom{-}$2.94 $\pm$     0.02   $\pm$ 0.02      \\
\end{tabular}
\label{FinalRes}
\end{center}
\end{table}

\begin{table}[!h]
\caption{\small Signal yield ($N_{\mbox{\footnotesize{sig}}}$) and results
for the S-wave parameters in each bin of $m(K^+\pi^-)$ mass, showing statistical and systematic uncertainties.
Only the physical solution is shown for $\Ds-\delta_0$.}
\begin{center}
\begin{tabular}{cccc}
$m(K^+\pi^-)$ $[\mevcc]$   & $N_{\mbox{\footnotesize{sig}}}$ & Parameter                       &  value \\
\hline

826 -- 861               &       6\;456 $\pm$ 69 & $\Fs$  &  0.115 $\pm$ 0.007 $\pm$ 0.020          \\
                        &                                       &  $\Ds - \delta_0$[rad] & 3.09 $\pm$ 0.10 $\pm$ 0.08          \\
\hline
861 -- 896               &      24\;418 $\pm$ 80 &        $\Fs$ &  0.049 $\pm$ 0.004 $\pm$ 0.007              \\
                        &                                       &  $\Ds - \delta_0$[rad] & 2.66 $\pm$ 0.06 $\pm$ 0.06         \\
\hline
896 -- 931               &       23\;036 $\pm$ 77 & $\Fs$  & 0.052 $\pm$ 0.006 $\pm$ 0.009            \\
                        &                                       &       $\Ds - \delta_0$[rad] & 1.94 $\pm$ 0.03 $\pm$ 0.09             \\
\hline
931 -- 966               &       7\;383 $\pm$ 64 & $\Fs$  & 0.105 $\pm$ 0.014 $\pm$ 0.008            \\
                        &                                        & $\Ds -
                        \delta_0$[rad]   & 1.53 $\pm$ 0.03 $\pm$ 0.11\\
\end{tabular}
\end{center}
\label{tab:simFit}
\end{table}

\begin{table}[!h]
\caption{\small Correlation matrix for the four-bin fit.}
\begin{center}
\footnotesize{
\begin{tabular}{c|cccccccccccc} 
& $|A_{\parallel}|^{2}$ & $|A_{\perp}|^{2}$ & $\delta_{\parallel}$ &
$\delta_{\perp}$ & $F_{\mathrm S}(1)$ & $\delta_{\mathrm S}(1)$ & 
$F_{\mathrm S}(2)$ &
$\delta_{\mathrm S}(2)$ & $F_{\mathrm S}(3)$ & $\delta_{\mathrm S}(3)$ &
$F_{\mathrm S}(4)$ &
$\delta_{\mathrm S}(4)$\\ \hline
$|A_{\parallel}|^{2}$ & 1.00  & -0.70 & 0.12  & 0.04  &
0.02  & -0.02 & 0.10  & -0.08 & 0.14  & -0.07 & 0.10  & 0.03  \\
$|A_{\perp}|^{2}$ &       & 1.00  & -0.14 & -0.01 & 0.01  &
0.02  & -0.09 & 0.12  & -0.19 & 0.15  & -0.15 & -0.01 \\
$\delta_{\parallel}$ &       &       & 1.00  & 0.64  & -0.01 & 0.06 & -0.06 &
0.10  & -0.07
& 0.12  & -0.02 & 0.07  \\
$ \delta_{\perp}$ &       &       &       & 1.00  & -0.03 &
0.14  & -0.16 & 0.21  & -0.17 & 0.18  & -0.09 & 0.05  \\
$ F_{\mathrm S}(1)$ &       &       &       &       & 1.00
& -0.24 & 0.01  & -0.01 & -     & - & -     & -     \\
$ \delta_{\mathrm S}(1)$ &       &       &       &       & & 1.00  & -0.03 &
0.03
& -0.03 & 0.03  & -0.02 & -     \\
$  F_{\mathrm S}(2)$ &       &       &       &       &       &       & 1.00  &
-0.76
& 0.05  & -0.04 & 0.03  & -     \\
$ \delta_{\mathrm S}(2)$ &       &       &       &       &        &       &
& 1.00  & -0.06 & 0.05  & -0.04 & 0.01  \\
$ F_{\mathrm S}(3)$ &       &       &       &       &       &       &       &
& 1.00  & -0.59 & 0.04  & -     \\
$ \delta_{\mathrm S}(3)$ &       &       &       &
&       &       &       &       &       & 1.00  & -0.04 & 0.01  \\
$F_{\mathrm S}(4)$ &       &       &       &       &
&       &       &       &       &       &   1.00  & 0.19  \\
$\delta_{\mathrm S}(4)$ &       &       &       &       &
&       &       &       &       &       &          & 1.00  \\
\end{tabular}
}
\end{center}
\label{tab:corr}
\end{table}

The results of separate fits to 30\;896 $\pm$ 95 $B^0\to J/\psi K^{*0}$ and 30\;442 $\pm$ 92 \mbox{$\overline{B}^0\to J/\psi\overline K^{*0}$} background
subtracted candidates are shown in Table \ref{directCP}, along with the direct \CP asymmeties. Only the P-wave amplitudes are
allowed to vary in the fit; the S-wave parameters in each $m(K^+\pi^-)$ bin are fixed to the 
values determined with the combined fit. 
The fit allows for a difference between the angular acceptance due to charge
asymmetries in the detector. The systematic uncertainties are calculated
similarly as described in Sec.~\ref{sec:systematics}; the uncertainty due to the angular acceptance
partially cancels in the direct \CP asymmetry calculation.  The $B^0$ and
$\overline{B}^0$ fit results are consistent within uncertainties, with the
largest
difference being approximately $2$ standard deviations in $\Aperpsq$. There is
no evidence for BSM
contributions to direct \CP violation at the current level of
precision.  

\begin{table}[!h]
\caption{\small Results from fits to the $B^0\to J/\psi K^{*0}$ and 
$\overline{B}^0\to J/\psi\overline K^{*0}$ background subtracted candidates 
and the direct \CP asymmetries
$\frac{\overline{X}-X}{\overline{X}+X}$, where $X$ represents the
parameter in question. The uncertainties are statistical for the amplitudes 
and phases and both statistical and systematic for the direct \CP measurements.}
\begin{center}
\begin{tabular}{ccccc}
           Parameter &     Value for $B^0$ & Value for $\overline{B}^0$ &
           $B^0$-$\overline{B}^0$ asymmetry	   \\  \hline
   $\Aparasq$ &        0.230 $\pm$    0.005     &   0.225 $\pm$    0.005     		&			$-$0.011 $\pm$ 0.016  $\pm$ 0.005 \\
       $\Aperpsq$ &         0.194 $\pm$    0.005  &  0.207 $\pm$     0.005            &			0.032 $\pm$ 0.018  $\pm$ 0.003	\\
  $\Dpara$ [rad] &         $-$2.94 $\pm$     0.03      &   $-$2.92 $\pm$     0.03         &					0.003 $\pm$ 0.007  $\pm$ 0.002	\\
      $\Dperp$ [rad] &       $\phantom{-}$2.94 $\pm$     0.02 &  $\phantom{-}$2.96 $\pm$      0.02            &				0.003 $\pm$ 0.005  $\pm$ 0.001\\
        \end{tabular}
\label{directCP}
\end{center}
\end{table}

\begin{table}[!h]
\footnotesize
\caption{\small Comparison of the LHCb results assuming no S-wave component
with results from previous experiments. The uncertainties are statistical and systematic, respectively.}
\begin{center}
\begin{tabular}{c|cccc}
& LHCb (no S-wave)& BaBar 2007  \cite{Collaboration:2007kx}               & Belle 2005 \cite{PhysRevLett.95.091601}     & CDF 2005 \cite{cdf}          \\
\hline
$\Aparasq$      		& $ 0.220 \pm 0.004 \pm 0.003 $	& $0.211 \pm 0.010 \pm 0.006 $          & $0.231 \pm 0.012 \pm 0.008$ &$0.211 \pm 0.012 \pm 0.009 $    \\
$\Aperpsq$            	&     $0.210 \pm 0.004 \pm 0.004$  & $0.233 \pm 0.010 \pm 0.005$           & $0.195 \pm 0.012 \pm 0.008 $ &$0.220 \pm 0.015 \pm 0.012$ \\
$\Dpara$ [rad]               	&    $-2.98 \pm 0.03 \pm 0.01$ 		& $-2.93 \pm 0.08 \pm 0.04$             & $-2.887 \pm 0.090 \pm 0.008$&$-2.97 \pm 0.08 \pm 0.03$    \\
$\Dperp$ [rad]                	&    $\phantom{-}2.97 \pm 0.02 \pm 0.02$   		& $\phantom{-}2.91 \pm 0.05 \pm 0.03$              & $\phantom{-}2.938\pm 0.064 \pm 0.010$ &  $\phantom{-}2.97 \pm 0.06 \pm 0.01$ \\
\end{tabular}
\label{tab:comp}
\end{center}
\end{table}

In previous analyses of the $B^0\to J/\psi K^{*}{^0}$ polarisation amplitudes
and phases fits have been performed using a single bin
in $m(K^+\pi^-)$ and no S-wave component has been included.  
To allow comparison with recent results, the fit is repeated in a single $m(K^+\pi^-)$ bin with the
S-wave component set to zero.
The results are summarised in Table~\ref{tab:comp} and
are consistent with the previous results, and are more accurate by a factor of 2
to 3. BaBar has also resolved the two-fold ambiguity in the strong
phases~\cite{Aubert:2005uq, Aubert:2001prd} but has not reported S-wave fractions in separate bins.


\section{Conclusion}
\label{sec:conclusion}

A full angular analysis of the decay $B^0\to J/\psi K{^*}{^0}$ has been performed.
The polarisation amplitudes and their strong phases are measured using data, corresponding to 
an integrated luminosity of $1.0\fb^{-1}$, collected in $pp$ collisions at a centre-of-mass
energy of $7 ~\mathrm{TeV}$ with the LHCb detector. 
The results are consistent with previous measurements and 
confirm the theoretical predictions mentioned in Sec.~\ref{sec:Introduction}. The ambiguity 
in the strong phases is resolved by measuring the relative S- and P-wave
phases in bins of the $K^+\pi^-$ invariant mass. 
No significant direct \CP asymmetry is observed.


\section*{Acknowledgements}
\label{sec:Acknowledgements}

\noindent We express our gratitude to our colleagues in the CERN
accelerator departments for the excellent performance of the LHC. We
thank the technical and administrative staff at the LHCb
institutes. We acknowledge support from CERN and from the national
agencies: CAPES, CNPq, FAPERJ and FINEP (Brazil); NSFC (China);
CNRS/IN2P3 and Region Auvergne (France); BMBF, DFG, HGF and MPG
(Germany); SFI (Ireland); INFN (Italy); FOM and NWO (The Netherlands);
SCSR (Poland); MEN/IFA (Romania); MinES, Rosatom, RFBR and NRC
``Kurchatov Institute'' (Russia); MinECo, XuntaGal and GENCAT (Spain);
SNSF and SER (Switzerland); NAS Ukraine (Ukraine); STFC (United
Kingdom); NSF (USA). We also acknowledge the support received from the
ERC under FP7. The Tier1 computing centres are supported by IN2P3
(France), KIT and BMBF (Germany), INFN (Italy), NWO and SURF (The
Netherlands), PIC (Spain), GridPP (United Kingdom). We are thankful
for the computing resources put at our disposal by Yandex LLC
(Russia), as well as to the communities behind the multiple open
source software packages that we depend on.

\bibliographystyle{LHCb}
\bibliography{main}

\end{document}